\documentclass[11pt,epsf]{article}
\usepackage[a4paper, margin=1.2in]{geometry} % Adjust margins as needed

\usepackage[colorlinks=true, linkcolor=blue, citecolor=blue, urlcolor=blue]{hyperref}
\usepackage{amsmath}
\usepackage{amsfonts}
\usepackage{amssymb}
\usepackage{booktabs}
\usepackage{needspace}
\usepackage{cleveref}

\usepackage{xfrac}
\usepackage{slashed}
\usepackage{hhline}
\usepackage{tikz}
\usetikzlibrary{shapes, arrows, fit, calc, positioning, chains, quotes}

\usetikzlibrary{arrows.meta,decorations.markings,shapes.geometric}
\usetikzlibrary{bending} 
\tikzset{
  arc arrow/.style args={%
    to pos #1 with length #2}{
    decoration={
      markings,
      mark=at position 0 with {\pgfextra{%
          \pgfmathsetmacro{\tmpArrowTime}{#2/(\pgfdecoratedpathlength)}
          \xdef\tmpArrowTime{\tmpArrowTime}}},
      mark=at position {#1-\tmpArrowTime} with {\coordinate(@1);},
      mark=at position {#1-2*\tmpArrowTime/3} with {\coordinate(@2);},
      mark=at position {#1-\tmpArrowTime/3} with {\coordinate(@3);},
      mark=at position {#1} with {\coordinate(@4);
        \draw[-{Latex[length=#2,bend]}]       
        (@1) .. controls (@2) and (@3) .. (@4);},
    },
    postaction=decorate,
  }
}

\usepackage{graphicx}
\usepackage{wrapfig}
\usepackage{subfig}

\usepackage[frozencache=true, cachedir=./]{minted}  % [cache=false]  % frozencache, cachedir=./.cache/   % finalizecache   % --shell-escape  % frozencache=true, cachedir=./
\makeatletter
\usepackage{etoolbox}
\AtBeginEnvironment{Verbatim}{
   \def\PYGvs@tok@err {\color{black} \def\PYGvs@bc##1{\strut ##1}}
}
\makeatother

\usepackage{lineno}
\modulolinenumbers[5]

\usepackage[utf8]{inputenc}
\DeclareUnicodeCharacter{27F9}{\ensuremath{\Longrightarrow}}
\DeclareUnicodeCharacter{2209}{\ensuremath{\notin}}
\DeclareUnicodeCharacter{266D}{\ensuremath{\flat}}
\DeclareUnicodeCharacter{2208}{\ensuremath{\in}}
\DeclareUnicodeCharacter{00E9}{\'{e}}
\DeclareUnicodeCharacter{2264}{\ensuremath{\leq}}
\DeclareUnicodeCharacter{2265}{\ensuremath{\geq}}
\DeclareUnicodeCharacter{2200}{\ensuremath{\forall}}
\DeclareUnicodeCharacter{221E}{\ensuremath{\infty}}
\DeclareUnicodeCharacter{221D}{\ensuremath{\propto}}
\DeclareUnicodeCharacter{210F}{\ensuremath{\hbar}}
\DeclareUnicodeCharacter{2020}{\ensuremath{\dagger}}
\DeclareUnicodeCharacter{03B1}{\ensuremath{\alpha}}
\DeclareUnicodeCharacter{03B2}{\ensuremath{\beta}}
\DeclareUnicodeCharacter{0393}{\ensuremath{\Gamma}}
\DeclareUnicodeCharacter{03B3}{\ensuremath{\gamma}}
\DeclareUnicodeCharacter{03B8}{\ensuremath{\theta}}
\DeclareUnicodeCharacter{21D2}{\ensuremath{\Rightarrow}}
\DeclareUnicodeCharacter{0394}{\ensuremath{\Delta}}
\DeclareUnicodeCharacter{03B4}{\ensuremath{\delta}}
\DeclareUnicodeCharacter{03C1}{\ensuremath{\rho}}
\DeclareUnicodeCharacter{03BE}{\ensuremath{\xi}}
\DeclareUnicodeCharacter{03C0}{\ensuremath{\pi}}
\DeclareUnicodeCharacter{03A0}{\ensuremath{\Pi}}
\DeclareUnicodeCharacter{03BC}{\ensuremath{\mu}}
\DeclareUnicodeCharacter{03BD}{\ensuremath{\nu}}
\DeclareUnicodeCharacter{222B}{\ensuremath{\int}}
\DeclareUnicodeCharacter{2211}{\ensuremath{\Sigma}}
\DeclareUnicodeCharacter{03C3}{\ensuremath{\sigma}}
\DeclareUnicodeCharacter{03C4}{\ensuremath{\tau}}
\DeclareUnicodeCharacter{03BB}{\ensuremath{\lambda}}
\DeclareUnicodeCharacter{03B7}{\ensuremath{\eta}}
\DeclareUnicodeCharacter{03A6}{\ensuremath{\Phi}}
\DeclareUnicodeCharacter{03D5}{\ensuremath{\phi}}
\DeclareUnicodeCharacter{03A8}{\ensuremath{\Psi}}
\DeclareUnicodeCharacter{03C8}{\ensuremath{\psi}}
\DeclareUnicodeCharacter{03F5}{\ensuremath{\epsilon}}
\DeclareUnicodeCharacter{2202}{\ensuremath{\partial}}
\DeclareUnicodeCharacter{220F}{\ensuremath{\prod}}
\DeclareUnicodeCharacter{03C9}{\ensuremath{\omega}}
\DeclareUnicodeCharacter{2260}{\ensuremath{\neq}}
\DeclareUnicodeCharacter{22C5}{\ensuremath{\cdot}}
\DeclareUnicodeCharacter{27E8}{\ensuremath{\langle}}
\DeclareUnicodeCharacter{27E9}{\ensuremath{\rangle}}
\DeclareUnicodeCharacter{2192}{\ensuremath{\rightarrow}}
\DeclareUnicodeCharacter{00B2}{\ensuremath{{}^2}}
\DeclareUnicodeCharacter{221A}{\ensuremath{\sqrt{}}}
\DeclareUnicodeCharacter{223C}{\ensuremath{\sim}}
\DeclareUnicodeCharacter{226A}{\ensuremath{\ll}}
\DeclareUnicodeCharacter{00D7}{\ensuremath{\times}}
\DeclareUnicodeCharacter{27EA}{\ensuremath{\langle\!\langle}}
\DeclareUnicodeCharacter{27EB}{\ensuremath{\rangle\!\rangle}}

\begin{document}

\title{
%\hfill ~\\[-30mm]
%\phantom{h} \hfill\mbox{\small IPPP/25/xx}
%\\[1cm]
%\vspace{10mm}
\textsc{Linac}: linear algebra with CUDA over finite fields
}

\date{}
\author{
Giuseppe De Laurentis$^{1\,}$\footnote{E-mail:
  \texttt{giuseppe.delaurentis@ed.ac.uk}},
Jack Franklin$^{2,\,3\,}$\footnote{E-mail:
  \texttt{jack.d.franklin@durham.ac.uk}}
\\[9mm]
{\small\it $^1$ Higgs Centre for Theoretical Physics, University of Edinburgh, } \\
{\small\it Edinburgh, EH9 3FD, United Kingdom}\\[3mm]
{\small\it $^2$ Institute for Particle Physics Phenomenology, Department of Physics, } \\
{\small\it University of Durham, Durham DH1 3LE, United Kingdom}\\[3mm]
{\small\it $^3$ Institute of Computing for Climate Science, University of Cambridge, } \\
{\small\it Wilberforce Road, Cambridge, CB3 0WA, United Kingdom}\\[3mm]
}
\maketitle

\begin{abstract}
  \noindent
Solving linear systems of polynomial equations is a ubiquitous problem
in both mathematics and physics. The standard approach, Gaussian
elimination, scales cubically with system size and often constitutes a
computational bottleneck. The algorithm's inherent parallelism makes
it well-suited for modern computing architectures, namely graphics
processing units (GPUs), which offer significantly higher throughput
than CPUs.  Additionally, the use of finite fields---integers modulo a
prime---in place of floating-point arithmetic offers a scalable
solution to the issue of numerical precision loss, which becomes
increasingly problematic at large system sizes. With \textsc{Linac},
we present a high-performance, open-source, parallel implementation of
Gaussian elimination over finite fields and floating-point
arithmetic. This tool has been developed for applications to analytic
reconstruction of scattering amplitudes in quantum field theory.
\end{abstract}
\thispagestyle{empty}
% \vfill
% \newpage
\setcounter{page}{1}

\newpage

\noindent\rule{\linewidth}{1pt}
\setcounter{tocdepth}{2}
\vspace{-2em}
\tableofcontents
\vspace{0.5em}
\noindent\rule{\linewidth}{1pt}

\section{Introduction}

Much of modern research in high-energy physics—spanning experimental
analyses, phenomenological simulations, and theoretical
computations—is driven by the balance between the increasing scope of
the calculations we aim to perform and the computational resources
available to realise them. Progress in algorithms is closely tied to
the efficiency of their implementation and to the performance
characteristics of the underlying hardware. Off-loading highly
parallelisable and computationally intensive operations from CPUs to
general-purpose GPUs has become an essential strategy to meet the
ever-growing computational demands across multiple areas of
high-energy physics \cite{HEPSoftwareFoundation:2017ggl,
  HSFPhysicsEventGeneratorWG:2020gxw}.

Some topics that have already seen benefits from GPU acceleration
include Monte Carlo phase-space integration \cite{Kanzaki:2010ym},
event reweighting \cite{Roiser:2025bsk}, parton showers
\cite{Seymour:2025fpu}, tree-level amplitude generation
\cite{Bothmann:2021nch, Cruz-Martinez:2025kwa, Valassi:2025gmq},
Feynman integrals and loop amplitudes \cite{Smirnov:2015mct,
  Borowka:2017idc, Borowka:2018goh, Heinrich:2021dbf,
  Heinrich:2023til}. The growing prevalence of GPUs in large-scale HPC
systems \cite{bib:summit95}, together with rapid advances in GPU
architectures and device memory, further strengthens the case for
GPU-based high-precision computations. Building on these developments,
the present work extends GPU acceleration to exact linear algebra over
finite fields and $p\kern0.2mm$-adic numbers, tools that have come to
play a central role in perturbative quantum field theory (QFT)
computations.

With \textsc{Linac}, we present an efficient, GPU-accelerated solution
to the ubiquitous problem of dense Gaussian elimination. The
implementation enables row reduction of arbitrary matrices, provided
their dense representation fits within the available device memory
(VRAM). The practical upper bound on (square) matrix dimension can be
estimated as\footnote{We often quote sizes for square matrices, as
this provides a single representative number and corresponds to the
case of full-rank linear systems. This is not a limitation of the
software: \textsc{Linac} supports rectangular matrices of arbitrary
shape.}
\begin{equation}
N_{\text{max}} \simeq \sqrt{\frac{\text{VRAM [bytes]}}{4}} =
\sqrt{ \frac{\text{VRAM [GB]}}{4} \times 10^{9} } \, ,
\end{equation}
for 32-bit number types, using decimal GB; for binary GiB replace
$10^9$ with $2^{30}$. On current high-end GPUs with 40 GB of VRAM,
this corresponds to systems of order $10^5 \times 10^5$ elements, and
beyond for cutting-edge hardware with more memory. As a preview of
performance on an NVIDIA A100 we observe the following timings,
\begin{align}
  \text{time}(N = 10^4) &\approx 3.6 \; \text{s} \; , \\
  \text{time}(N = 10^5) &\approx 45 \; \text{min} \; .
\end{align}
More details on performance are given in section \ref{sec:benchmark}.
The memory limitation stems from the need to avoid costly data
transfers between the host (CPU) and the device (GPU) during
intermediate stages of the computation. The present work focuses on
dense linear systems and assumes a dense in-memory representation of
the system matrix. Sparse matrices are supported insofar as they can
be accommodated within a dense representation. In this case, sparsity
can still lead to performance gains, since zero entries increase the
frequency with which certain branches of the algorithm are bypassed
(e.g.~due to vanishing pivots or early rank deficiency). Moreover,
even when a problem is initially sparse, intermediate stages of the
computation often become dense, in which case transitioning to a dense
algorithm is both natural and advantageous.

Another central consideration in Gaussian elimination is numerical
stability. An initial version of the package presented here, limited
to floating-point arithmetic, was presented in the appendix of the PhD
thesis of one of the authors~\cite{DeLaurentis:2020xar}. Beyond
performance, a key motivation for a custom GPU implementation was the
ability to explicitly control numerical thresholds---in particular, to
decide which small quantities are treated as zero. The current
implementation extends this earlier version to multiple arithmetic
domains,
\begin{equation}
\text{supported fields } \mathbb{F} \text{:} \; \{ \mathbb{R}, \mathbb{C} \, ,
\mathbb{F}_p \; \text{and} \; \mathbb{Q}_p \; (\text{leading digit})\} \, .
\end{equation}
The inclusion of finite fields, $\mathbb{F}_p$, represents an
important development for high-precision QFT
computations. Finite-field arithmetic was first introduced in this
context for\linebreak integration-by-parts (IBP)
reductions~\cite{vonManteuffel:2014ixa} and subsequently for analytic
reconstruction via interpolation
techniques~\cite{Peraro:2016wsq}. Over the past decade, finite-field
methods have been integrated into many standard tools used in
high-energy physics computations. These include IBP reduction
frameworks such as \textsc{Fire}~\cite{Smirnov:2019qkx} and
\textsc{Kira}~\cite{Maierhofer:2017gsa, Klappert:2020nbg}, as well as
dedicated reconstruction and interpolation libraries such as
\textsc{FireFly}~\cite{Klappert:2019emp, Klappert:2020aqs},
\textsc{FiniteFlow}~\cite{Peraro:2019svx}, and the optimisation tool
\textsc{Ratracer}~\cite{Magerya:2022hvj}. Finite-field techniques have
also been successfully employed in generalized-unitarity amplitude
computations \textsc{Caravel}~\cite{Abreu:2020xvt}. More
general-purpose implementations of finite-field arithmetic, such as
\textsc{FiniteFieldSolve}~\cite{Mangan:2023eeb} and
\textsc{Galois}~\cite{Hostetter_Galois_2020}, have also become
available.

GPU-accelerated linear algebra dates back to the early
2000s~\cite{inproceedingsGaloppo}, and has seen significant
developments in floating-point applications. However, its use in
combination with exact arithmetic over finite fields in high-energy
physics remains comparatively unexplored. The present work addresses
this gap by providing a GPU-based implementation of Gaussian
elimination tailored to such settings. Crucially, the finite-field
reduction mode implemented in \textsc{Linac} extends naturally to the
leading digit of $p\kern0.2mm$-adic numbers~\cite{DeLaurentis:2022otd,
  Chawdhry:2023yyx}. This enables the solution of linear systems
arising in singular limits, where quantities scale with negative
powers of $p$, within the same computational framework.

In this context, it is natural to ask to what extent Gaussian
elimination can be avoided altogether in favour of interpolation-based
approaches. While in some applications Gaussian elimination can be
circumvented—for example, by exploiting the structured nature of
Vandermonde systems \cite{Klappert:2019emp, Abreu:2021asb} or through
Newton or Thiele interpolation—these approaches rely on the ability to
cast the problem into a compatible form. This has not been shown to be
always possible. In particular, to the best of our knowledge,
interpolation algorithms are not generally formulated for computations
over quotient rings, where variables are subject to algebraic
constraints. Such quotient structures arise naturally in many physical
contexts \cite{DeLaurentis:2022otd, Cederwall:2025ywy}, for example
when working with redundant kinematic variables subject to constraints
such as momentum conservation and Schouten identities. In
spinor-helicity variables, these redundancies appear already at low
multiplicity due to the two-component nature of spinors, while for
four-momenta they arise starting at five points, where linear
dependence among four-vectors sets in. In these settings, Gaussian
elimination remains the most general and robust method for solving
linear systems, regardless of the underlying field or ring structure.

This paper is organised as follows. In \cref{sec:library}, we provide
a user-level introduction to the \textsc{Linac} Python library,
covering installation and its core functionality. In
\cref{sec:algorithms}, we discuss programming on GPUs and describe in
detail our implementation of Gaussian elimination on such
architectures. We also address the construction of linear systems from
polynomial data and the optimisations employed. In
\cref{sec:benchmark}, we present performance results on different
hardware platforms, demonstrating the clear advantage over CPU-based
implementations. In \cref{sec:applications}, we discuss higher-level
applications of the core functionality of the library, including rank
factorisation, null-space computation, solving linear systems,
vector-space manipulations, and ansatz fitting. Finally, in
\cref{sec:conclusion}, we summarise the work presented here and
outline possible future directions.

\section{Quick Start}\label{sec:library}

\subsection{Dependencies}

The \textsc{Linac} library requires a minimal set of core Python packages, chiefly:
\begin{itemize}
  \item \texttt{numpy} \cite{harris2020array};
\end{itemize}
with mild dependencies on:
\begin{itemize}
  \item \texttt{pyadic} (for the extended Euclidean algorithm and
    finite-field class \texttt{ModP});
  \item \texttt{syngular} (for the definition of \texttt{Field}).
\end{itemize}
Optional dependencies are grouped into extras:
\begin{itemize}
  \item \texttt{[cuda]}: GPU acceleration support via \texttt{pycuda},
    enabling the GPU-accelerated linear-algebra routines that form the
    core of this library;
  \item \texttt{[dev]}: development, testing and additional tools, including
    \texttt{mpmath}~\cite{mpmath},
    \texttt{galois}~\cite{Hostetter_Galois_2020},
    \texttt{diskcache}, \texttt{pytest}, \texttt{pytest-cov},
    and \texttt{flake8};
  \item \texttt{[full]}: includes all packages from both \texttt{[cuda]} and \texttt{[dev]}.
\end{itemize}
All Python dependencies are automatically installed by \texttt{pip}.

To enable GPU acceleration, the system must provide a functional CUDA
development environment.  Specifically, \texttt{pycuda} requires:

\begin{itemize}
  \item the \texttt{nvcc} compiler;
  \item compatible NVIDIA drivers with access to a CUDA-enabled GPU.
\end{itemize}
The availability of these can be verified using \texttt{nvcc
  --version} and \texttt{nvidia-smi}. These must be installed
\emph{before} installing the \texttt{[cuda]} or \texttt{[full]}
extras.

A minimal working setup is demonstrated in the GitHub continuous
integration workflow \texttt{ci\_test.yml}, which runs on a
self-hosted Docker container described in \cref{app:github-actions-runner}.

\subsection{Installation}

The \textsc{Linac} library is fully open source. It can be found on GitHub at
\begin{center}
\href{https://github.com/GDeLaurentis/linac}{\texttt{github.com/GDeLaurentis/linac}}
\end{center}
and on Zenodo \cite{giuseppe_de_laurentis_2026_20327732}.
Its latest stable release can be installed from PyPI
via
\vspace{-3mm}
\begin{minted}[escapeinside=||, mathescape=true, linenos=False, numbersep=5pt, gobble=2, frame=lines, framesep=2mm, breaklines, breakautoindent=false, breakindent=-12.5pt, ]{bash}
  pip install --upgrade linac
\end{minted}
\vspace{-1.5mm}
Optional extras can be installed by specifying them in square brackets, for example:
\vspace{-3mm}
\begin{minted}[escapeinside=||, mathescape=true, linenos=False, numbersep=5pt, gobble=2, frame=lines, framesep=2mm, breaklines, breakautoindent=false, breakindent=-12.5pt, ]{bash}
  pip install linac[cuda]    # for GPU acceleration
  pip install linac[dev]     # for development and additional tools
  pip install linac[full]    # for all optional features
\end{minted}
\vspace{-1.5mm}
Alternatively, for development purposes or to use the latest source, clone the repository and install locally in editable mode:
\vspace{-3mm}
\begin{minted}[escapeinside=||, mathescape=true, linenos=False, numbersep=5pt, gobble=2, frame=lines, framesep=2mm, breaklines, breakautoindent=false, breakindent=-12.5pt, ]{bash}
  git clone git@github.com:GDeLaurentis/linac.git
  pip install -e path/to/linac/[extra]
\end{minted}
\vspace{-1.5mm}
where \texttt{extra} can be any of \texttt{cuda}, \texttt{dev}, or \texttt{full} as needed.

\subsection{Basic Usage}

The main user-facing entry point of \textsc{Linac} is the function
\vspace{-3mm}
\begin{minted}[escapeinside=||,mathescape=true,frame=lines,framesep=2mm,breaklines]{python}
cuda_row_reduce(matrix, field_characteristic=0, verbose=False)
\end{minted}
\vspace{-1.5mm}
which performs Gaussian elimination on the GPU and returns a
row-reduced echelon form (RREF) stored in the original matrix shape,
with zero rows possibly appearing at the bottom when the rank is
deficient.\footnote{A convenience routine
\texttt{drop\_bottom\_zero\_rows} is provided in the module
\texttt{linear\_algebra\_tools} to remove these rows when a strictly
minimal representation of the RREF is required.}

The argument \texttt{matrix} is a two-dimensional \texttt{numpy.ndarray}
containing the input data. The parameter \texttt{field\_characteristic}
specifies the arithmetic domain: setting it to zero selects real or
complex floating-point arithmetic, while a positive integer \(p\)
selects exact arithmetic modulo \(p\). The optional flag
\texttt{verbose} enables diagnostic output.

The characteristic of a field is the smallest positive integer \(n\)
such that
\[
\underbrace{1 + 1 + \cdots + 1}_{n\ \text{times}} = 0 .
\]
If no such \(n\) exists, the field has characteristic zero; in
particular, \(\mathbb{Q}_p\) has characteristic zero. In
\textsc{Linac}, however, the parameter \texttt{field\_characteristic}
is also used to select reduction over the residue field associated
with \(p\kern0.2mm\)-adic arithmetic. Concretely, setting
\texttt{field\_characteristic=p} performs arithmetic in
\(\mathbb{F}_p \cong \mathbb{Z}_p / p\mathbb{Z}_p\), corresponding to
the leading \(p\kern0.2mm\)-adic digit. This convention provides a
uniform interface for finite-field and \(p\kern0.2mm\)-adic workflows.

Internally, \textsc{Linac} selects numerical backends automatically:
real matrices use \texttt{float64}, complex matrices use
\texttt{complex128}, and modular arithmetic uses \texttt{uint32} or
\texttt{uint64} depending on the size of \(p\). Specifically,
\texttt{uint32} is used for \(p \le 2^{32}-1\) and \texttt{uint64} for
\(2^{32}-1 < p \le 2^{64}-1\). Larger primes are not supported.
Optional parameters allow forcing a specific backend for testing and
benchmarking.

Random test matrices may be generated using \texttt{numpy.random.rand}
for floating-point arithmetic and \texttt{numpy.random.randint} for
finite-field computations. When working in \(p\kern0.2mm\)-adic
leading-digit mode, inputs should be supplied after stripping the
leading power of \(p\). In cases where the leading
\(p\kern0.2mm\)-adic valuation is not uniform, care must be taken to
ensure that the resulting system is meaningful; in practice, this
requires entries starting at higher powers of \(p\) to appear as zeros
at the level of the leading digit.

\medskip

A second core functionality of \textsc{Linac} is the construction of
dense matrices representing polynomial linear systems, via
\vspace{-3mm}
\begin{minted}[escapeinside=||, mathescape=true, linenos=False, numbersep=5pt, gobble=2, frame=lines, framesep=2mm, breaklines, breakautoindent=false, breakindent=-12.5pt, ]{python}
cuda_load_matrix(bases, lindices, field_characteristic=0, verbose=False)
\end{minted}
\vspace{-1.5mm}
or, at a higher level,
\vspace{-3mm}
\begin{minted}[escapeinside=||, mathescape=true, linenos=False, numbersep=5pt, gobble=2, frame=lines, framesep=2mm, breaklines, breakautoindent=false, breakindent=-12.5pt, ]{python}
load_matrices(prefactors, ansatze, points, use_cuda=True, verbose=False)
\end{minted}
\vspace{-1.5mm}
Both routines evaluate monomials at numerical points and assemble the
corresponding matrices, either on the GPU or on the CPU.

The low-level function \texttt{cuda\_load\_matrix} assumes that the
numerical values of a set of ``basis variables'' have already been
computed. These are passed as a two-dimensional array \texttt{bases}
of shape \((n_{\text{rows}}, n_{\text{basis}})\), where each row
corresponds to one numerical sampling point and each column to one
basis variable. The monomial structure is encoded by an integer array
\texttt{lindices} of shape \((n_{\text{cols}}, d)\): each row of
\texttt{lindices} lists \(d\) column indices into \texttt{bases} whose
product defines one monomial.\footnote{The CUDA kernel assumes a
uniform monomial degree \(d\); the helper \texttt{load\_matrices} pads
shorter monomials with the dummy variable \(1\).} The returned matrix
\(A\) has shape \((n_{\text{rows}}, n_{\text{cols}})\) with entries
\[
  A_{r c} \;=\; \prod_{t=1}^{d} \texttt{bases}_{r,\;\texttt{lindices}_{c t}} \, ,
\]
so that each column corresponds to one monomial evaluated over all
sampling points.

The wrapper \texttt{load\_matrices} provides a more convenient
interface for typical workflows. It accepts one or more prefactors
(e.g.~inverse denominators) and ansätze (i.e.~lists of monomials),
together with vectorised evaluation points. These may be provided
either as a list-like container carrying a \texttt{field} attribute
and capable of evaluating the monomials (e.g.~\texttt{RingPoints} from
\texttt{syngular}), or as a dictionary mapping variable names to arrays
of precomputed values and containing a \texttt{field} entry. The
routine constructs the \texttt{bases} and \texttt{lindices} arrays
internally and returns a list of matrices, one per system. When
\texttt{use\_cuda=True}, matrix construction is off-loaded to
\texttt{cuda\_load\_matrix}; otherwise, a CPU backend is used, with
optional finite-field acceleration via vectorised \texttt{ufunc}
implementations provided by \textsc{Galois} when available. Arithmetic
domains are selected consistently with \texttt{cuda\_row\_reduce},
supporting real and complex floating-point arithmetic as well as exact
arithmetic over \(\mathbb{F}_p\) (and leading-digit
\(p\kern0.2mm\)-adic mode).

In practice, these routines are most easily accessed indirectly
through higher-level code. For instance, we use them via the
\texttt{Terms.\_\_call\_\_} interface in \textsc{Antares}
\cite{giuseppe_de_laurentis_2026_18894183}, where \texttt{Terms}
represent rational functions with numerator ansätze and the call
argument is a collection of numerical points in the relevant
(quotient) ring. Nevertheless, they are exposed here as standalone
utilities for users who wish to assemble linear systems directly.

\medskip

In \cref{sec:applications}, we provide additional functionalities
derived from these core ones. Namely, the construction of rank
factorisations, null-spaces, solutions of linear systems, abstractions
through \texttt{Python} classes for the description of vector spaces,
and a preview of derived functionality in \textsc{Antares} for fitting
ans\"atze in the context of functional reconstruction.

\section{Algorithms}\label{sec:algorithms}

The main capability of \textsc{Linac} is row reduction to reduced
row-echelon form via parallel Gaussian elimination on NVIDIA graphics
cards using CUDA. While the algorithm is well known and part of many
undergraduate curricula, its parallelisation on GPUs is not
necessarily straightforward, especially in conjunction with finite
field arithmetic.

\subsection{Fundamentals of CUDA programming}

Programming on GPUs is fundamentally different from programming on
CPUs. Code execution is intrinsically parallelised across hundreds or
thousands of \emph{cores}, but high performance is not guaranteed. To
achieve good performance, one must carefully structure computation and
data access, keeping in mind the hardware's parallel architecture. For
instance, threads executing in parallel should ideally follow the same
instruction sequence; otherwise, executions stall and performance may
suffer. We briefly review here some fundamental concepts of CUDA
programming, and refer the reader to ref.~\cite{nvidia-cuda-2025} for
further details.

A first distinction has to be drawn between the \emph{host} and the
\emph{device}. The host refers to the CPU side of the computation. It
is responsible for preparing input data, launching GPU kernels, and
collecting the output. The device refers to the GPU side of the
computation. It handles the computationally intensive, massively
parallelised tasks. We use GPU kernel calls from the host to
synchronise the execution on the device side.

Data transfer between host and device is subject to significant
latency and relatively low bandwidth compared to on-device memory
access. Therefore, it is preferable to minimise communication between
host and device. In our implementation we limit communication to the
beginning and end of the computation. In this way, the overhead is
negligible for sufficiently intensive computations. This choice,
however, implies that the host remains unaware of intermediate stages
of the computation and therefore cannot direct or adjust the procedure
accordingly.

In CUDA, kernel execution on the device side is organised
hierarchically in a 2D \emph{grid} of 3D \emph{blocks}, each
containing execution \emph{threads}. Threads within the same block can
communicate via fast on-chip \emph{shared memory}, while communication
across blocks must occur through the slower global memory. At the
hardware level, threads are executed in groups of 32, known as
\emph{warps}. All threads in a warp must follow the same instruction
sequence to be executed concurrently. Warp divergence, caused when
threads within a warp take different control flow paths (e.g. due to
different branches of an \texttt{if} statement), cause performance
loss.

Device functions are launched from the host via kernel calls, which
coordinate execution on the GPU. Blocks are scheduled in an
unspecified order by the GPU hardware. Therefore, explicit
synchronisation across blocks typically relies on host-side kernel
launch. Within a block, threads can be synchronised with an explicit
\texttt{\_\_syncthreads()} call. This should, however, only be used
when strictly necessary as it may degrade performance.

Memory access patterns have a significant impact on GPU
performance. Since threads within a warp are executed simultaneously,
it is most efficient for consecutive threads to access consecutive
memory locations (coalesced access), allowing the GPU to fetch data
efficiently in fewer transactions. In contrast, strided or irregular
memory accesses result in additional memory transactions,
significantly reducing throughput. For this reason, access patterns of
the data in device memory are a crucial aspect of high-performance
linear algebra. Moreover, since all memory is effectively stored as a
one-dimensional sequence, rather than, for example, two-dimensional
for a matrix, careful handling of the indexing is required to ensure
the correct elements are read and written.

Another important performance consideration is \emph{stalls}. They
occur when a thread cannot proceed because it is waiting for data, for
a dependent instruction to complete, or for other threads to reach a
synchronisation point. GPUs hide some of this latency by switching to
executing other ready warps on the same multiprocessor; however, if
too many threads are stalled simultaneously, performance eventually
suffers. Minimising stalls requires balancing of the available
resources to maximise parallelism, though depending on the nature of
the algorithm, some stalls may be unavoidable.

In the next subsection, we describe our implementation of Gaussian
elimination in \textsc{Linac}, highlighting the practical impact of
the concepts discussed above.

\subsection{Gaussian elimination on GPUs}\label{sec:GaussianElimination}

\paragraph{Chosen algorithm.} %
We implement partially pivoted row reduction to reduced echelon form
in CUDA. Since the Gaussian elimination algorithm is well known, we
will focus our attention on the GPU implementation and finite field
arithmetic. Recall the difference between the three typical types of
pivoting: 1) partial pivoting looks for the maximum within a column,
resulting in permutations of equations but not variables; 2) rook
pivoting looks for the first non-zero entry in the full sub-block left
to row reduce, stopping as soon as a non-zero entry is found; 3) total
pivoting performs a full search of the sub-block matrix left to row
reduce for the largest entry. The latter two types can result in
permutations of variables, besides equations. They are more
computationally expensive and potentially more stable for
floating-point reduction, but do not provide any benefit for reduction
over finite field. We implement all three pivoting strategies in the
CPU function \texttt{row\_reduce}, while the GPU function
\texttt{cuda\_row\_reduce} uses partial pivoting.

\paragraph{Number types.} %
After some initial bookkeeping to decide on a correct number type,
\texttt{float64}, \texttt{complex128}, \texttt{int32}, or
\texttt{int64}, we copy the matrix to be row reduced from the host to
the device. The copy happens with padding, yielding an effective
number of columns which may not be exactly the true number of columns
of the matrix. This is done so that vectorised memory access
(i.e.~accessing multiple matrix elements at once, up to 4 for
\texttt{int32}) is guaranteed not to overflow into the next
row. Recall that the memory layout on the device is flat: row and
column numbers need to be combined into a single index. The copy is
done once at the beginning and from that moment on, until the end of
the computation, the device will not be aware of the state of the
matrix being row reduced. This imposes some constraints on how the
kernels are designed and called. For instance, we cannot decide to
call two different kernels depending on whether a non-zero pivot can
be found in a given column.

\paragraph{Meta-programming and pre-compiler.} %
The \texttt{row\_reduce.cu} kernel is compiled on the fly by our
custom function \texttt{cuda\_set\_vars\_and\_get\_funcs} which wraps
\texttt{pycuda}, see \cref{app:metaprog} for more details. This
function exposes to the host all functions after comment mark \texttt{
  // GLOBAL FUNCTIONS}, providing \texttt{Python} functions with a
prefix \texttt{Cuda} followed by the name of the kernel defined in the
\texttt{.cu} file. The compilation overhead is negligible and
compiling on the fly provides us with a simple way to achieve one of
the most drastic optimisations: having a compile-time constant field
characteristic (i.e.~the value of the prime for the modular
arithmetic). This allows the compiler to optimise the modulo reduction
operator \texttt{\%} in terms of bit shifts and other basic
operations, resulting in a significant speedup. We observe the speedup
obtained in this way to be comparable to casting the whole computation
to Montgomery form, which, however, is significantly more
cumbersome. Number types are set by the pre-compiler to match those
coming from the host. Through C++ \texttt{using} statements, we can
make much of the algorithm run interchangeably among number types,
while the remaining parts that depend on the number type are wrapped
in pre-compiler if statements, such as \texttt{\#if
  FIELD\_CHARACTERISTIC~==~0}.

\paragraph{Rescaling} %
The first two kernel calls happen only with approximate
(floating-point type) number types. These are one-off kernels that are
called once at the beginning of the computation. Their objective is to
mitigate any potential scale differences present in the matrix, which
can adversely affect floating-point stability. The first kernel
\texttt{CudaSetRowScales} is a reduction algorithm which finds the
maximum of each row. The second kernel is \texttt{CudaRescaleRows},
which divides each row by the corresponding maximum found in the
previous call.

\paragraph{Looping} %
All following kernel calls happen in a loop, whereby this sequence of
kernels is called as many times as the number of columns of the
matrix. This is important to ensure a row reduced echelon form
irrespective of the shape of the input matrix. An optional
\texttt{verbose} keyword argument prints the progress in the
calculation, reported as the current iteration number over the
total. The rate of progress through the loop is not constant: the
calculation speeds up as the sub-matrix left over to row reduce
shrinks.

\paragraph{Pivoting} %
The first kernels in the loop implement partial pivoting. First, two
kernels locate the greatest entry in a row by locating the maximum
within sub-blocks of the column \texttt{CudaThreadsReduceToMaxIndex}
and then finding the maximum among these via
\texttt{CudaBlocksReduceToMaxIndex}. For finite fields, finding any
non-zero entry would suffice, and we may choose to call a separate
kernel in that case --- however, this step is so subdominant overall
that we have not found this optimisation worth it yet. After having
located the maximum within a column, the kernel
\texttt{CudaSwitchRows} swaps the two rows in memory on the device. At
this point, the maximum may still be a zero.

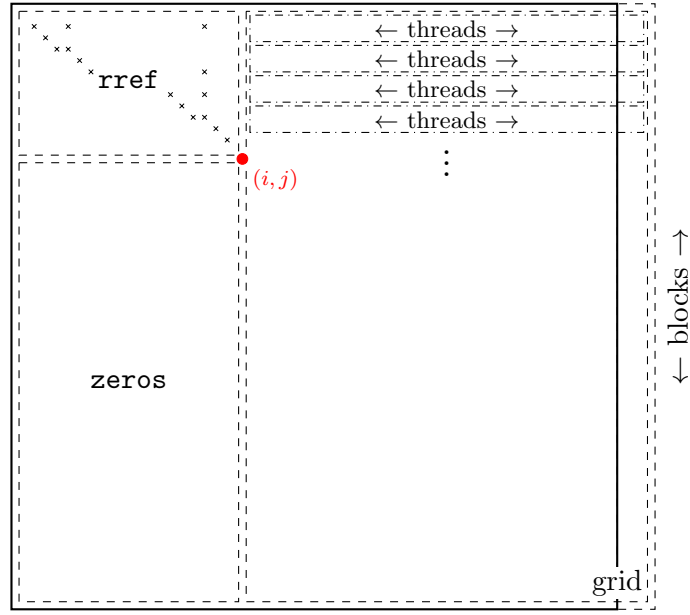
\begin{figure}[t!]
\centering
\begin{tikzpicture}[scale=1]

  % matrix
  \draw[thick] (0, 0) rectangle (8, -8);
  \draw[dashed] (8, 0) rectangle (8.5, -8);

  % Add pivot and non-pivot dots inside rref box
  % Rows: roughly evenly spaced vertically, columns: spaced horizontally
  \foreach \x/\y in {0.3/-0.3, 0.45/-0.45, 0.6/-0.6, 0.75/-0.6, 0.75/-0.3, 0.9/-0.75, 1.05/-0.9, 2.1/-1.2, 2.25/-1.35, 2.4/-1.5, 2.55/-1.5, 2.55/-1.2, 2.55/-0.9, 2.55/-0.3,      2.7/-1.65, 2.85/-1.8 } {
    \draw[black, line width=0.3pt]
      (\x-0.03, \y-0.03) -- (\x+0.03, \y+0.03)
      (\x-0.03, \y+0.03) -- (\x+0.03, \y-0.03);
  }
  
  % rref sub-matrix
  \draw[dashed] (0.1, -0.1) rectangle (3, -2);
  \node[fill=white, inner sep=1pt] at (1.55, -1) {\normalsize \texttt{rref}};
  
  % zeros sub-matrix
  \draw[dashed] (0.1, -2.1) rectangle (3, -7.9);
  \node at (1.55, -5) {\normalsize \texttt{zeros}};

  \node[fill=white, inner sep=1pt] at (8, -7.65) {\normalsize grid };

  % blocks and threads
  \draw[dashed] (3.1, -0.1) rectangle (8.4, -7.9);
  \node[rotate=90] at (8.8, -4) {\normalsize $\leftarrow$ blocks $\rightarrow$};
  
  % \draw[dashdotted] (3.15, -0.15) rectangle (8.35, -0.5);
  % \node at (5.75, -0.3) {\small $\leftarrow$ threads $\rightarrow$};

  % repeated thread bands (4 visible)
  \foreach \y in {0.15, 0.55, 0.95, 1.35 } {
    \draw[dashdotted] (3.15, -\y) rectangle (8.35, -\y - 0.35);
    \node at (5.75, -\y - 0.175) {\footnotesize $\leftarrow$ threads $\rightarrow$};
  }

  % vertical ellipsis
  \node at (5.75, -2) {\Large$\vdots$};
  
  % Mark (i,j)
  \filldraw[red] (3.05, -2.05) circle (2pt) node[below right] {\scriptsize$(i,j)$};

\end{tikzpicture}
\caption{Matrix layout for CUDA \texttt{RowReduce} kernel: thread blocks operate to the right of column $(j)$.\label{fig:rowred}}
\end{figure}

\paragraph{Conditional Rescale} %
The following kernel, \texttt{CudaCompareHeadToTolerance}, checks
\linebreak whether the head of the row is zero, or near zero in case
of floating-point numbers, and saves the result in a global device
boolean. By head we mean the first entry in the sub-block left to row
reduce. In figure~\ref{fig:rowred}, the head is the entry at
coordinates $(i, j)$. In the $j^\text{th}$ iteration, the column index
is guaranteed to be $j$, while the row index may be lower in case a
non-trivial null-space was found until this point. If the head is not
zero, the following kernel call, \texttt{CudaConditionalRescaleRow},
rescales it to a 1. This kernel needs to be called in any case, as the
host is not aware of the result of the comparison.

\paragraph{Row reduce} %
The main kernel, which dominates the total execution time, is
\texttt{RowReduce}. It is invoked through the wrapper
\texttt{ConditionalRowReduce} to correctly handle the case of a zero
head. Figure~\ref{fig:rowred} schematically illustrates the grid
structure of this kernel call. We employ a one-dimensional grid of
one-dimensional thread blocks. Each block operates on a single matrix
row, say row~$k$, and performs two elementary arithmetic operations:
it multiplies the corresponding entry in row~$j$ by the head of
row~$k$ and subtracts the result from the entry in row~$k$. Depending
on the underlying data type, entries are fetched in 128-bit chunks, so
each thread may update multiple matrix elements per iteration.

When the logical row length exceeds 1024~entries—the maximum number of
threads per block—it is not possible to assign one thread to each
matrix column directly. In this situation we ``fold'' the row: each
thread processes several disjoint segments of the row in a strided
fashion. Crucially, threads within a block always access consecutive
memory locations at each iteration of the update loop, ensuring
coalesced memory accesses, while the striding occurs only along the
per-thread iteration direction, which has no adverse effect on
bandwidth. Folding therefore allows us to support arbitrarily long
rows without modifying the kernel structure or sacrificing
memory-access efficiency.

After all updates for row~$k$ have been completed, a
\texttt{\_\_syncthreads()} ensures that every thread has finished
reading the original head element before it is overwritten. The head
of row~$k$ is then set to zero.

\paragraph{Increment counters} %
The last kernel call within the loop is to increment the column index
$j$ and, if the pivot was non-zero, the row index $i$.

\subsection{Building linear systems on GPUs}\label{sec:systembuilding}

A common application of Gaussian elimination is to solve for
undetermined numerical coefficients $c_i$ in equations of the form
\begin{equation}\label{eq:ansatz}
  c_1 \, m_{1}(\underline X) +  \dots + c_n \, m_{n}(\underline X) = f(\underline X)
\end{equation}
where the $m_i$'s are monomials in the variables $\underline X$, and
$f$ is a function to be evaluated. The LHS of eq.~\ref{eq:ansatz} is
often referred to as an ansatz for $f$. To solve for the coefficients,
one can build a system of equations at random points $\underline
X_{1},\dots, \underline X_{n}$
\begin{equation}
  \begin{pmatrix}
    m_{1} (\underline X_1) & \dots & m_{n} (\underline X_1) \\
    \vdots & \ddots & \vdots \\
    m_{1} (\underline X_n) & \dots & m_{n} (\underline X_n) 
  \end{pmatrix} \cdot \begin{pmatrix} c_1 \\ \vdots \\ c_n \end{pmatrix} = \begin{pmatrix} f(\underline X_1) \\ \vdots \\ f(\underline X_n) \end{pmatrix} \, ,
\end{equation}
yielding $n$ equations in $n$ unknowns. Constructing such a system can
itself be computationally intensive and yet easily parallelisable by
evaluating the same monomials in parallel at different numerical
points. The computational complexity is
$\mathcal{O}(n^2\text{deg}(p))$, assuming uniform degree monomials,
which is consistent with a plot given in the next section.

\begin{figure}[t!]
\centering
\begin{tikzpicture}[scale=1]

  % matrix
  \draw[thick] (0, 0) rectangle (8, -4);

  % blocks and threads
  \draw[dashed] (0.1, -0.1) rectangle (7.9, -3.9);
  \node[rotate=90] at (8.4, -2) {\normalsize $\leftarrow$ blocks $\rightarrow$};

  % repeated thread bands (4 visible)
  \foreach \y in {0.15, 0.55, 0.95, 1.35 } {
    \draw[dashdotted] (0.15, -\y) rectangle (7.85, -\y - 0.35);
    \node at (3.75, -\y - 0.175) {\footnotesize $\leftarrow$ threads $\rightarrow$};
  }

  % vertical ellipsis
  \node at (3.75, -2) {\Large$\vdots$};

\end{tikzpicture}
\caption{Matrix layout for CUDA \texttt{LoadMatrix} kernel.\label{fig:matrixbuild}}
\end{figure}
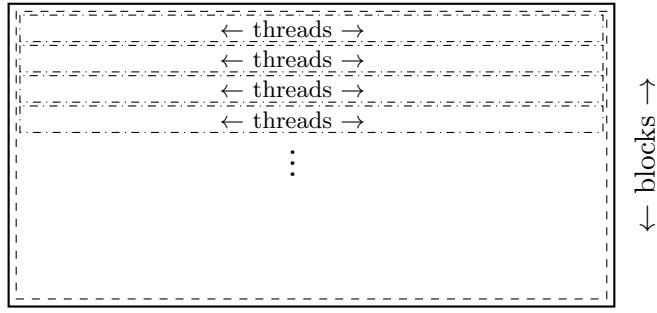

We implement this construction in \texttt{cuda\_load\_matrix}, which
calls \texttt{matrix\_loader.cu}. The numerical values $\underline
X_{1},\dots, \underline X_{n}$ are first prepared on the CPU and then
passed to the GPU, while the monomials \(m_i\) are encoded as integer
vectors specifying the variables entering each monomial. The GPU
kernel then evaluates the monomials by multiplying the variables as
specified by these integer vectors in parallel across rows,
corresponding to distinct numerical points, as shown in
\cref{fig:matrixbuild}.

Each block requires the same set of $\underline X_{i}$, therefore the
first step is to load all the data into shared memory (recall this is
local to the block). After a \texttt{\_\_syncthreads()} call ensuring
that all the data has been fetched, each thread works on a different
monomial, computing the multiplication reduction. If the number of
columns exceeds 1024, the maximum number of threads in a block, then
each thread works on multiple monomials, as many as necessary to
complete the computation.

\subsection{Optimisations}

Several low-level optimisations were implemented in order to maximise
throughput on modern GPU architectures. The most significant
improvement comes from promoting the field characteristic \(p\) to a
compile-time constant through CUDA meta-programming (see also
appendix~\ref{app:metaprog}). Concretely, the CUDA source is generated
dynamically from Python with definitions such as
\vspace{-3mm}
\begin{minted}[frame=lines,framesep=2mm,breaklines]{c}
#define FIELD_CHARACTERISTIC ...
#define NBR_ROWS ...
#define NBR_COLUMNS ...
\end{minted}
\vspace{-1.5mm}
inserted before compilation. This allows the compiler to optimise the
modular reduction operator \texttt{\%} into specialised arithmetic
operations tailored to the specific prime, such as bit shifts and
other low-level integer operations. Together with warp-aligned thread
configurations, where thread-block sizes are rounded to multiples of
32,
\vspace{-3mm}
\begin{minted}[frame=lines,framesep=2mm,breaklines]{python}
round_to_multiple_of(..., 32)
\end{minted}
\vspace{-1.5mm}
this yields a conservative performance improvement of roughly
\(50\%\).

Another important optimisation concerns memory access patterns. Matrix
entries are processed in vectorised chunks using packed data types
such as \texttt{uint4}, \texttt{ulonglong2}, and \texttt{double2},
depending on the arithmetic domain. For instance, \texttt{uint4}
groups four 32-bit unsigned integers into a single 128-bit access and
is therefore used for finite fields with
\(p \leq 2^{32}-1\). This allows each thread to load and update
multiple matrix elements at once through coalesced memory transactions,
improving effective memory-bandwidth utilisation and providing an
additional \(10\%\)--\(20\%\) speedup in practice.

A further optimisation comes from the fact that portions of the
elimination algorithm are conditionally skipped whenever the relevant
entries already vanish. This becomes increasingly beneficial in the
presence of rank deficiency or partial sparsity, where large parts of
the computation may be bypassed automatically without requiring a
specialised sparse-matrix implementation.

As shown in section~\ref{sec:benchmark}, the resulting kernels are
predominantly limited by global-memory bandwidth rather than by
arithmetic throughput. Further low-level optimisations of the present
algorithm are therefore expected to yield only marginal gains; more
substantial improvements would require increasing the arithmetic
intensity or otherwise reducing global-memory traffic, which is
challenging given the nature of Gaussian elimination.

\section{Benchmarks}\label{sec:benchmark}

\subsection{Gaussian elimination performance}

\begin{figure}[t!]
    \centering
    \includegraphics[width=0.95\textwidth]{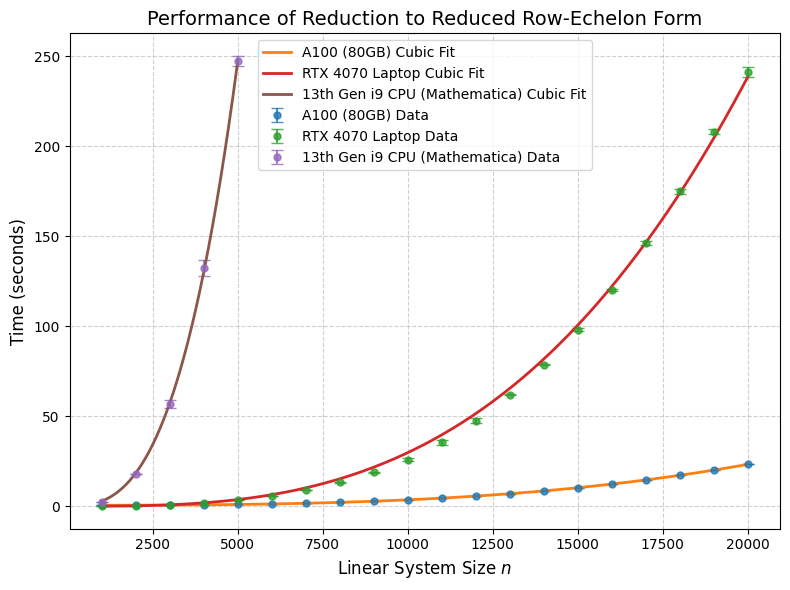}
    \captionsetup{justification=centering}
    \caption{Cubic fit of the data points illustrating the performance of Gaussian elimination to reduced row-echelon form.}
    \label{fig:cubic_fit}
\end{figure}

We study the performance of the Gaussian elimination algorithm
implemented in \textsc{Linac} by running it over 100\% dense, fully
random square matrices in $\mathbb{F}_{2^{31}-1}$, with size 1k to
20k, in steps of 1k. We fit the resulting data points to a cubic, as
shown in fig.~\ref{fig:cubic_fit}. We do this for three typical
hardware use cases, namely:
\begin{itemize}
  \item an NVIDIA workstation GPU, A100 with 80GB of VRAM (blue data,
    orange fit);
  \item an NVIDIA laptop GPU, RTX 4070 with 8GB of VRAM (green data,
    red fit);
  \item an Intel (laptop) CPU, $13^{\text{th}}$ Gen i9 CPU, single
    thread (purple data, brown fit).
\end{itemize}
The CPU code is the function \texttt{RowReduce} from
\texttt{Mathematica} 13. We observe virtually identical performance by
running the \texttt{Python} algorithm \texttt{row\_reduce} implemented
in \texttt{linac/row\_reduce.py} when the input matrix is taken to be
a \texttt{numpy} extension type (Galois field) from the
\textsc{Galois} library \cite{Hostetter_Galois_2020}. In this case,
the keyword argument \texttt{prime} should be left at the default
value, \texttt{None}, so that arithmetic is delegated to the extension
type. This suggests that the observed performance is close to that of
the underlying compiled implementation, possibly with a small
overhead. As expected, the same algorithm using native \texttt{Python}
integers runs slower (by about 50\%), although this comparison is not
entirely fair, since integers in \texttt{Python~3} support
arbitrary-precision arithmetic rather than being restricted to
fixed-size \texttt{int32}/\texttt{int64} representations.

The GPU code is the function \texttt{cuda\_row\_reduce} from
\texttt{linac/pycuda\_row\_reduce.py}, which compiles on the fly and
runs the CUDA script \texttt{linac/row\_reduce.cu}. The implementation
is described in section~\ref{sec:GaussianElimination}. The code
running on the two GPUs is identical.

\begin{table}[t!]
\centering \resizebox{\linewidth}{!}{%
\begin{tabular}{lccccc}
\hline
Name & $a_3$ & $a_2$ & $a_1$ & $a_0$ & Goodness of Fit ($R^2$) \\
\hline
A100 (80GB) & $2.77 \times 10^{-12}$ & $3.58 \times 10^{-17}$ & $3.00 \times 10^{-5}$ & $0.50$ & $0.99994$ \\
\hline
RTX 4070 Laptop & $2.98 \times 10^{-11}$ & $4.16 \times 10^{-24}$ & $5.69 \times 10^{-22}$ & $9.65 \times 10^{-23}$ & $0.99891$ \\
\hline
13th Gen i9 CPU & $ 1.76 \times 10^{-9}$ & $1.13 \times 10^{-6}$ & $1.92 \times 10^{-15}$ & $7.61 \times 10^{-12}$ & $0.99989$ \\
\hline
\end{tabular}%
}
\captionsetup{justification=centering}
\caption{
  Gaussian elimination fit parameters to $y = a_3 x^3 + a_2
  x^2 + a_1 x + a_0$, \\ where $y$ is the execution time and $x$ is
  system size.  }
\label{tab:fitparameterscubic}
\end{table}

We fit a cubic to the data points, obtaining the fit parameters shown
in table~\ref{tab:fitparameterscubic}. As shown by the $R^2$ in the
last column, the fit is excellent, especially on the workstation GPU,
whose performance is less affected by fluctuations in clock frequency
due to thermal or power limitations. Except for a constant half-second
overhead ($a_0$) observed in the workstation case, the timing is
heavily dominated by the coefficient of the cubic term. By taking
ratios of the cubic-term coefficients ($a_3$), we observe that the GPU
algorithm is roughly equivalent to \textbf{$\boldsymbol{60}$ CPU cores
  on a laptop}, and \textbf{$\boldsymbol{600}$ CPU cores on a
  workstation}, demonstrating a clear and significant advantage of the
GPU implementation.\footnote{These figures assume that multi-threaded
CPU performance scales linearly with the single-threaded result, and
that cluster CPUs have a similar clock frequency to the laptop
CPU. Both assumptions are usually optimistic, so the comparison is
conservative.}

Our CPU benchmark is in line with the reported performance of
\textsc{FiniteFieldSolve} \cite[Table B.6]{Mangan:2023eeb}: evaluating
the CPU fit parameters at 2000 we obtain 18.5 seconds, while the
reported time in the latter reference is 40\% of 41 seconds, i.e.~16.5
seconds.  It seems clear that the \texttt{Solve} function in
Mathematica has significant overheads, given that the native
\texttt{RowReduce} implementation gives our reported timing. By
comparison, on the same laptop, our GPU implementation runs in 0.23
seconds.
  
\subsubsection{Profiling the row-reduction kernel}\label{sec:profiling}

Since the vast majority of the total execution time (typically in
excess of $90\%$) is spent in the \texttt{RowReduce} kernel, it is
natural to ask what the dominant bottlenecks are and to what extent
further performance gains are achievable. To address this question, we
profiled the kernel using \textsc{Nsight Compute} on the devices whose
performance we quoted above: a laptop NVIDIA RTX~4070 GPU and a
workstation-class NVIDIA A100 GPU. The terminal command reads
\vspace{-3mm}
\begin{minted}[escapeinside=||, mathescape=true, linenos=False, numbersep=5pt, gobble=2, frame=lines, framesep=2mm, breaklines, breakautoindent=false, breakindent=-12.5pt, ]{bash}
  ncu --kernel-name ConditionalRowReduce --launch-skip 50
      --launch-count 1 --set full "python" script_name.py
\end{minted}
\vspace{-1.5mm}
where the script runs \texttt{cuda\_row\_reduce} over a random
$8k\times 8k$ matrix.

The profiling results indicate that the kernel operates in a
memory-bandwidth--limited regime. In particular, we observe sustained
global-memory throughput at approximately $70$--$90\%$ of the
theoretical peak bandwidth of the device, corresponding to $\sim
200$~GB/s on the RTX~4070 and $\sim 1.6$~TB/s on the A100. At the same
time, the achieved compute (SM) throughput remains moderate, around
$60\%$ on the RTX~4070 and $43\%$ on the A100, with a large fraction
($30$--$50\%$) of cycles spent stalled on memory dependencies, despite
near-maximal achieved warp occupancy ($97$--$98\%$). This indicates
that data movement, rather than arithmetic throughput, is the dominant
performance-limiting factor.

This behaviour is expected given the structure of Gaussian
elimination: each elimination step streams data from both the pivot
row and the target row, performs a small fixed number of arithmetic
operations per matrix element, and writes the updated row back to
global memory. The resulting arithmetic intensity is therefore low,
and improvements in raw compute capability alone do not lead to
substantial speedups.

Overall, this analysis suggests that the current implementation
operates close to the hardware-imposed memory-bandwidth limit, and
that any further performance gains would require algorithmic changes
that reduce global-memory traffic rather than low-level kernel
optimisations.

\subsection{Linear system construction performance}

We perform an analogous performance analysis for the construction of
linear systems in matrix form from polynomial equations. On the GPU,
this is handled by the host function \texttt{cuda\_load\_matrix},
which invokes the CUDA device code described in
section~\ref{sec:systembuilding}. On the CPU, we use vectorised
\texttt{numpy} operations provided by the \textsc{Galois} library
\cite{Hostetter_Galois_2020}, which approximate native \texttt{C++}
performance for finite-field arithmetic.

\begin{figure}[t!]
    \centering
    \includegraphics[width=0.93\textwidth]{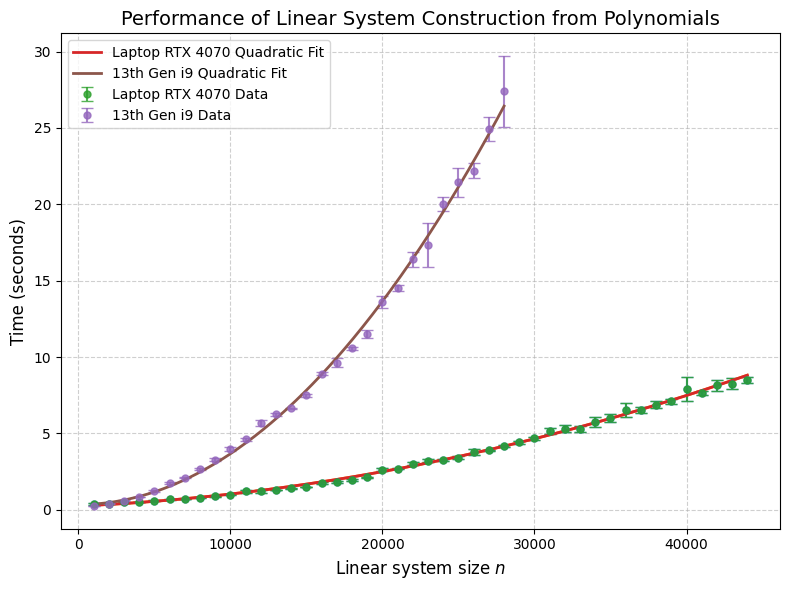}
    \captionsetup{justification=centering}
    \caption{Quadratic fit of the data points illustrating the performance
      of constructing an $n\times n$ matrix representation of a linear system
      from polynomial equations.}
    \label{fig:quadratic_fit}
\end{figure}

For benchmarking, we time the \texttt{Terms.\_\_call\_\_} method from
the \textsc{Antares} library, which wraps the above functionality via
the \texttt{load\_matrices} interface. The results are shown in
figure~\ref{fig:quadratic_fit} for the same laptop GPU (RTX~4070) and
CPU (13th Gen i9) considered in the previous section. We do not report
corresponding results for the workstation-class A100 GPU, since its
performance is essentially identical to that of the laptop GPU up to a
small constant overhead of order $\mathcal{O}(0.5\,\mathrm{s})$.
Indeed, on both devices the actual numerical work is already
subdominant, and the total runtime is dominated by Python-side
bookkeeping overhead. Even the one-time \textsc{pycuda} compilation
cost exceeds the execution time of the kernel itself.

The construction of the linear system proceeds by evaluating a set of
monomials at random numerical points. This operation scales linearly
with the number of matrix entries and hence quadratically with the
linear system size. We therefore fit the timing data to a quadratic
polynomial and find excellent agreement, as quantified by the $R^2$
values reported in table~\ref{tab:fitparametersquadratic}.

In contrast to the Gaussian elimination step, the speedup achieved on
the GPU is comparatively modest, corresponding to roughly the
performance of $\mathcal{O}(10)$ CPU cores. This is consistent with
the observation that the dominant cost in this stage arises from
Python bookkeeping rather than from arithmetic or memory throughput on
the device. Given that matrix construction is already strongly
subdominant in the overall runtime, we do not pursue further
optimisation of this component at present.

For small system sizes, we observe behaviour consistent with that
reported in Ref.~\cite[Table~B.6]{Mangan:2023eeb}, where the cost of
row reduction is comparable to that of matrix construction. For
example, for systems of size $2\times 10^3$, on the laptop GPU we find
timings of approximately $0.23\,\mathrm{s}$ for row reduction and
$0.40\,\mathrm{s}$ for construction. However, this is clearly an
artefact of the limited system size. As the dimension increases, the
asymptotic scaling rapidly separates the two contributions: for a size
of $2\times 10^4$, we measure approximately $241\,\mathrm{s}$ for
Gaussian elimination versus $2.6\,\mathrm{s}$ for matrix
construction. For sufficiently large systems, the total runtime is
therefore overwhelmingly determined by the efficiency of the
row-reduction kernel rather than by the cost of building the linear
system.

\begin{table}[t!]
  \centering
\resizebox{0.8\linewidth}{!}{%
\begin{tabular}{lcccc}
\hline
Name & $a_2$ & $a_1$ & $a_0$ & Goodness of Fit ($R^2$) \\
\hline
RTX 4070 Laptop & $3.42 \times 10^{-9}$ & $4.41 \times 10^{-5}$ & $2.44 \times 10^{-1}$ & $0.99717$ \\
\hline
13th Gen i9 CPU & $3.32 \times 10^{-8}$ & $1.01 \times 10^{-19}$ & $3.36 \times 10^{-1}$ & $0.99758$ \\
\hline
\end{tabular}%
}
\captionsetup{justification=centering}
\caption{
  Matrix construction fit parameters to $y =  a_2 x^2 + a_1 x + a_0$, \\
  where $y$ is the execution time and $x$ is system size.
}
\label{tab:fitparametersquadratic}
\end{table}

\section{Applications}\label{sec:applications}

Efficient Gaussian elimination has a wide range of applications in
scientific computing, from rank factorisation and the solution of
linear systems to more specialised problems involving vector spaces,
function spaces, and analytic reconstruction. Beyond the low-level
routines described in the previous sections, \textsc{Linac} also
provides higher-level interfaces built on top of row reduction.

We first describe two direct applications of the reduced row-echelon
form returned by \texttt{cuda\_row\_reduce}: the construction of rank
factorisations and null spaces, and the solution of linear systems. We
then discuss two vector-space abstractions: spaces generated by
explicit numerical column vectors, represented by
\texttt{ColumnVectorSpace}, and spaces spanned by functions, whose
linear relations are inferred from numerical evaluations, represented
by \texttt{VectorSpaceOfFunctions}. Finally, we explain how some of
these ideas enter ansatz fitting in analytic reconstruction.

\subsection{Rank factorisation and null spaces}

A direct application of row reduction is the construction of a rank
factorisation and of a basis for the null space of a matrix. Let
\(A\) be an \(m\times n\) matrix over a field \(\mathbb{F}\). In
\textsc{Linac}, one may first compute its row-reduced echelon form and
remove the zero rows,
\Needspace{6\baselineskip}
\vspace{-3mm}
\begin{minted}[escapeinside=||,mathescape=true,frame=lines,framesep=2mm,breaklines]{python}
rref = cuda_row_reduce(A, field_characteristic=p)
R = drop_bottom_zero_rows(rref)
\end{minted}
\vspace{-1.5mm}
or equivalently use \texttt{row\_reduce} on the CPU. If
\(r=\operatorname{rank}(A)\), then \(R\) has shape \(r\times n\), and
its rows form a basis for the row space of \(A\).

The pivot columns of \(R\), extracted for instance using
\Needspace{5\baselineskip}
\vspace{-3mm}
\begin{minted}[escapeinside=||,mathescape=true,frame=lines,framesep=2mm,breaklines]{python}
J = pivot_columns_from_row_reduced_echelon_form(R)
\end{minted}
\vspace{-1.5mm}
identify a set of linearly independent columns of the original matrix
\(A\). If \(C=A[:,J]\) is the submatrix of \(A\) formed from these
columns, then
\begin{equation}
  A = C R
\end{equation}
is a rank factorisation of \(A\). Indeed, the pivot columns of \(R\)
form the identity matrix, so each column of \(R\) gives the
coordinates of the corresponding column of \(A\) in the basis \(C\) of
the column space.

The same row-reduced form also determines the null space. Since row
operations preserve the solution space of the homogeneous system,
\begin{equation}
  A x = 0 ,
\end{equation}
one has \(\ker(A)=\ker(R)\). A canonical basis for this space can be
constructed as
\Needspace{5\baselineskip}
\vspace{-3mm}
\begin{minted}[escapeinside=||,mathescape=true,frame=lines,framesep=2mm,breaklines]{python}
K = canonical_kernel_from_row_reduced_echelon_form(R)
\end{minted}
\vspace{-1.5mm}
where the columns of \(K\) span the null space,
\begin{equation}
  A K = 0 .
\end{equation}
The construction uses the non-pivot columns of \(R\) as free variables
and fixes their coefficients to minus the identity matrix. This gives
a canonical representative of the null space determined by the pivot
structure of the RREF.

Thus, from a single row-reduction step, \textsc{Linac} provides the
rank of the matrix, bases for its row and column spaces, a rank
factorisation, and a canonical basis for its null space.

\subsection{Solving linear systems}

A second standard application of Gaussian elimination is the solution
of linear systems. Given a matrix \(A\) and a right-hand side \(b\),
one may solve
\begin{equation}
  A x = b
\end{equation}
by row reducing the augmented matrix
\begin{equation}
  M = (A\; b) .
\end{equation}
In the square full-rank case, the RREF of \(M\) has the form
\begin{equation}
  \big(I\; x\big),
\end{equation}
with $I$ the identity matrix, so the solution is read off from the
last column. In code, this corresponds schematically to
\Needspace{7\baselineskip}
\vspace{-3mm}
\begin{minted}[escapeinside=||,mathescape=true,frame=lines,framesep=2mm,breaklines]{python}
M = numpy.block([A, b.reshape(-1, 1)])
rref = cuda_row_reduce(M, field_characteristic=p)
x = rref[:, -1]
\end{minted}
\vspace{-1.5mm}
with the understanding that this direct extraction assumes that the
left block has reduced to the identity.

More generally, row reduction applies equally well when \(A\) is
rectangular or rank deficient. The system is consistent if and only if
the augmented column does not become a pivot column, equivalently if
\begin{equation}
  \operatorname{rank}(A)=\operatorname{rank}(A\;|\;b).
\end{equation}
If the augmented column is a pivot column, the RREF contains a row of
the form
\begin{equation}
  (0,\ldots,0\;|\;1),
\end{equation}
and the system has no solution. Otherwise, the pivot columns of \(A\)
identify the variables which are solved for, while the non-pivot
columns of \(A\) parametrise the free variables. In this case the RREF
describes an affine space of solutions. A particular solution is
obtained by setting the free variables to zero, while the remaining
solutions are obtained by adding elements of the null space of \(A\).
Thus the same RREF computation covers square full-rank systems,
overdetermined but consistent systems, and underdetermined systems
with non-unique solutions.

Ansatz fitting, discussed in section~\ref{sec:ansatz-fitting}, is a
direct instance of this construction. The unknown vector \(x\) is the
vector of coefficients in the ansatz, while the rows of \(A\) are
obtained by evaluating basis monomials or basis functions at numerical
sampling points.

\subsection{Vector spaces over number fields}

Given vectors \(v_i \in \mathbb{F}^m\), with coefficients in the same
field \(\mathbb{F}\), one can define vector spaces of the form
\begin{equation}
  V
  =
  \operatorname{span}_{\mathbb{F}}\{v_1,\ldots,v_n\}
  =
  \left\{
    \sum_{i=1}^{n} c_i v_i
    \;:\;
    c_i \in \mathbb{F}
  \right\}
  \subseteq \mathbb{F}^m .
\end{equation}
The class \texttt{ColumnVectorSpace} represents such spaces through
explicit matrix representatives. It is instantiated as
\Needspace{6\baselineskip}
\vspace{-3mm}
\begin{minted}[escapeinside=||,mathescape=true,frame=lines,framesep=2mm,breaklines]{python}
ColumnVectorSpace(matrix, prime=2147483647, use_galois=None,
                  use_gpu=None, verbose=False)
\end{minted}
\vspace{-1.5mm}
where \texttt{matrix} is a two-dimensional array whose columns are the
generators \(v_i\) of the vector space. The optional parameter
\texttt{prime} specifies the field: if \texttt{prime} is an integer,
arithmetic is performed over \(\mathbb{F}_p\), while
\texttt{prime=None} selects rational arithmetic over \(\mathbb{Q}\).

The optional parameters \texttt{use\_gpu} and \texttt{use\_galois}
select the computational backends. When left to \texttt{None}, the
default behaviour is chosen automatically according to the available
dependencies and the arithmetic domain. Boolean values provide explicit
overrides.

The class implements basic vector-space operations in terms of row
reduction. Containment of one space in another is tested by adjoining
the corresponding matrix representatives and checking whether the
pivot structure changes. Intersections are computed similarly. If
\(V\) and \(W\) are represented by matrices \(A\) and \(B\), whose
columns generate the two spaces, then
\[
  v \in V \cap W
\]
can be written as
\[
  v = A x = B y
\]
for some coefficient vectors \(x\) and \(y\). The intersection is
therefore obtained from the solutions of
\[
  (A \; -B)
  \begin{pmatrix}
    x \\ y
  \end{pmatrix}
  = 0 ,
\]
which are found by applying row reduction to the block matrix
\((A\; -B)\). The corresponding common vectors \(v=A x=B y\) then
provide a set of generators for \(V\cap W\).

As an example application, this class is used to represent linear
relations among residues of rational functions and to compute their
intersections. The latter is the core operation of the
rational-function basis-change algorithms used in
refs.~\cite{DeLaurentis:2023nss,DeLaurentis:2025dxw,DeLaurentis:2026brm}.

\subsection{Vector spaces over function fields}

A closely related problem arises when the objects whose span is to be
studied are functions rather than constant numerical vectors. Let
\(K\) denote a field of functions, for instance the field of fractions
of a polynomial quotient ring,
\begin{equation}
  K
  =
  \operatorname{Frac}\!\left(
    \mathbb{F}[\underline X]/
    \big\langle q_1(\underline X), \ldots, q_m(\underline X)
    \big\rangle
  \right),
\end{equation}
where the \(q_i\) are polynomials in the variables
\(\underline X=\{x_1,\ldots,x_n\}\). Given functions \(f_i\in K\),
and coefficients in \(\mathbb{F}\), one can define vector spaces of
the form
\begin{equation}
  V
  =
  \operatorname{span}_{\mathbb{F}}\{f_1,\ldots,f_n\}
  =
  \left\{
    \sum_{i=1}^{n} c_i f_i
    \;:\;
    c_i \in \mathbb{F}
  \right\}
  \subseteq K .
\end{equation}
The class \texttt{VectorSpaceOfFunctions} represents such spaces
through numerical evaluations of the functions. It is instantiated as
\Needspace{7\baselineskip}
\vspace{-3mm}
\begin{minted}[escapeinside=||,mathescape=true,frame=lines,framesep=2mm,breaklines]{python}
VectorSpaceOfFunctions(functions_evaluator, input_generator,
                       field=Field("finite field", 2**31 - 1, 1),
                       verbose=True, use_gpu=None,
                       iteration_start=20, iteration_step=20,
                       max_iteration=50, Cores=8)
\end{minted}
\vspace{-1.5mm}
where \texttt{functions\_evaluator} is a callable returning the values
of the functions \(f_i\) at a numerical point, typically represented
using the \texttt{tensor\_function} interface described in
\cref{app:tensor-functions}, and \texttt{input\_generator} is a
callable generating pseudo-random sampling points from an integer
seed. The parameter \texttt{field}\footnote{See
\texttt{syngular.field.Field}
\cite{giuseppe_de_laurentis_2026_18881385}.} specifies the field over
which the numerical evaluations and row reductions are performed;
finite fields are the intended use case. As above,
\texttt{use\_gpu=None} selects the default behaviour according to the
available CUDA dependencies, while Boolean values provide explicit
overrides.

The construction proceeds by evaluating the functions at a collection
of sampling points \(X_j\). This produces an evaluation matrix
\begin{equation}
  M_{j i} = f_i(X_j),
\end{equation}
whose columns correspond to the candidate functions and whose rows
correspond to numerical samples. Row reduction of this matrix identifies
a set of pivot columns, which are then used as a basis for the span.
Equivalently, $\mathbb{F}$-linear relations among the \(f_i\) are inferred from
linear relations among their numerical evaluations. The resulting pivot
columns are stored in the attribute \texttt{pivots}. They specify which
of the original functions form the chosen basis. The corresponding
reduced list of functions is stored in \texttt{basis\_functions}, so
that subsequent operations are performed using this smaller spanning
set rather than the original, possibly redundant, list of functions.

The number of sampling points is increased iteratively until the pivot
structure stabilises. The parameters \texttt{iteration\_start},
\texttt{iteration\_step}, and \texttt{max\_iteration} control this
process. Function evaluations may be parallelised over several CPU
cores through the parameter \texttt{Cores}, while the row reduction
itself can be off-loaded to the GPU.

This representation is useful when the functions are expensive or
impractical to manipulate symbolically, or not even known in symbolic
form, but can be evaluated efficiently at numerical points. The class
also implements containment tests for a function or another vector
space of functions by evaluating both on common sampling points and
checking whether adjoining the additional function or functions changes
the pivot structure. In addition, spaces can be enlarged by closing
them under a prescribed set of symmetry transformations acting on the
sampling points.

\subsection{Ansatz fitting}
\label{sec:ansatz-fitting}

As a final application, we briefly describe how \textsc{Linac} enters
the ansatz-fitting framework being developed in \textsc{Antares}
\cite{giuseppe_de_laurentis_2026_18894183}, focusing on the role
played by the linear-algebra routines presented in this work.

In analytic reconstruction problems, one often seeks to determine an
unknown, usually rational, function from numerical evaluations. After
suitable denominator information has been obtained or conjectured, the
remaining problem is to determine the coefficients of a numerator
ansatz. In schematic form, one writes
\begin{equation}
  f(X)
  =
  \sum_{\alpha} c_{\alpha}\, m_{\alpha}(X),
\end{equation}
or, more generally,
\begin{equation}
  f(X)
  =
  \sum_{r}
  \frac{1}{d_r(X)}
  \sum_{\alpha} c_{r,\alpha}\, m_{r,\alpha}(X),
\end{equation}
where the \(m_{\alpha}\) or \(m_{r,\alpha}\) are monomials in the
chosen variables, the \(d_r\) are known denominator factors, and the
\(c_{\alpha}\), \(c_{r,\alpha}\) are unknown coefficients in the
chosen field. Evaluating this ansatz at numerical points \(X_j\)
produces a dense linear system for the unknown coefficients,
\begin{equation}
  \sum_{\alpha} c_{\alpha}\, m_{\alpha}(X_j)
  =
  f(X_j) ,
\end{equation}
or the analogous system with multiple denominators.

The construction and solution of these systems is precisely the task
addressed by the low-level routines described in
section~\ref{sec:library}: \texttt{load\_matrices} assembles the
evaluation matrices, while \texttt{cuda\_row\_reduce} performs the row
reduction over finite fields. In \textsc{Antares}, the public
\texttt{Terms} class provides a representation of rational functions
with numerator ansätze and denominator data. Its
\texttt{\_\_call\_\_} interface evaluates such objects on collections
of numerical points and uses \textsc{Linac} to construct the
corresponding linear systems.

In realistic amplitude applications, however, ansatz fitting involves
additional layers beyond Gaussian elimination itself. These include the
construction of suitable ansätze, the management of denominator
information, the reuse of numerical samples, and caching strategies
for repeatedly fitting variations of closely related systems. The
higher-level fitting routines implementing this workflow are still
under active development and will be documented separately.

For this reason, \textsc{Linac} is best viewed as providing the
linear-algebra backend for a broader reconstruction pipeline rather
than as a complete ansatz-fitting framework by itself.

\section{Conclusions \& Outlook}\label{sec:conclusion}

We have presented an efficient GPU implementation of Gaussian
elimination to reduced row-echelon form using CUDA, achieving
substantial performance improvements over presently available
CPU-based approaches. The code interfaces with \texttt{pycuda} and
integrates seamlessly with \texttt{numpy}, the standard array library
in scientific computing. It is open source and available on both
\href{https://pypi.org/project/linac/}{\texttt{PyPI}} and
\href{https://github.com/GDeLaurentis/linac}{\texttt{GitHub}},
accompanied by continuous-integration infrastructure through
self-hosted GitHub Actions and by online Sphinx documentation.

The motivation for \textsc{Linac} comes from the growing role of exact
linear algebra over finite fields and \(p\kern0.2mm\)-adic numbers in
high-precision perturbative QFT computations. In this context, large
linear systems arise chiefly in integration-by-parts reduction and
finite-field reconstruction, allowing the determination of analytic
scattering amplitudes from numerical data. Although the benchmarks
presented here focus on dense systems arising in analytic
reconstruction, the row-reduction routines themselves are not tied to
this specific application and may be useful more broadly wherever
reasonably large dense linear systems over finite fields, or the
leading digits of \(p\kern0.2mm\)-adic numbers, need to be solved. The
implementation also retains support for real and complex
floating-point arithmetic, which may be useful in other applications.

We hope that making public a core component used in several recent
one- and two-loop amplitude computations will be useful to the broader
high-energy theory community, and may find applications beyond it.
Representative examples include analytic expressions for multi-leg QCD
amplitudes obtained by analysing numerical data produced with
\textsc{Caravel}~\cite{Abreu:2020xvt}, such as \(pp\rightarrow
jjj\)~\cite{DeLaurentis:2023nss,DeLaurentis:2023izi}, \(pp\rightarrow
Vjj\)~\cite{DeLaurentis:2025dxw}, and \(pp\rightarrow Hjj\)
\cite{DeLaurentis:2026brm}. Together with \textsc{lips}
\cite{giuseppe_de_laurentis_2026_20041968}, which provides kinematic
and spinor-helicity tools, \textsc{pyadic}
\cite{giuseppe_de_laurentis_2026_18881428}, which implements finite
fields, \(p\kern0.2mm\)-adic numbers, and lightweight interpolation
routines, and \textsc{syngular}
\cite{giuseppe_de_laurentis_2026_18881385}, which provides
algebraic-geometry functionality building on \textsc{Singular}
\cite{DGPS}, the \textsc{Linac} package introduced in this work forms
part of a broader numerical approach to analytic computation being
developed in \textsc{Antares}
\cite{giuseppe_de_laurentis_2026_18894183}. The aim is to replace,
where advantageous, symbolic manipulations in computer-algebra systems
by numerical finite-field and \(p\kern0.2mm\)-adic computations whose
output can nevertheless be analysed to recover exact analytic
expressions. This point of view is especially useful in polynomial
quotient rings, where algebraic relations among variables are imposed
from the outset. In such settings, symbolic manipulation can become
cumbersome, while numerical evaluation and linear algebra remain
comparatively efficient.

Highly efficient row reduction provides substantial headroom for
scaling analytic reconstruction to more complex systems. For example,
the largest systems solved in the recent \(pp\rightarrow Hjj\)
calculation of ref.~\cite{DeLaurentis:2026brm} had dimension of order
\(1.2\times 10^4\), well within the range where the GPU implementation
presented here remains fast, even on a laptop. At the same time,
improvements in reconstruction are complementary to improvements in
the generation of numerical data, for instance from
integration-by-parts reductions or numerical-unitarity calculations:
better reconstruction strategies reduce the number of sample points
needed, while faster sample generation reduces the cost of each
point. Finally, fast Gaussian elimination also enables automated
searches for improved representations of the reconstructed
expressions, such as changes of denominator structure leading to
lower-degree or otherwise simpler rational coefficients. A version of
this strategy was used in the \(pp\rightarrow Hjj\) calculation, where
such searches were incorporated as an automated simplification step.

There are several natural directions for future development. These
include adding the option to return an LUP factorisation in addition
to a rank decomposition; extending support to non-CUDA hardware, such
as AMD GPUs; exposing lower-level functionality through a native C++
API; and possibly developing an implementation capable of handling
very large but sparse matrices. The latter is particularly relevant
for direct applications to integration-by-parts reduction, where
sparse linear systems play a central role. Such an extension would
complement the dense row-reduction routines presented here and could
make GPU acceleration useful at an earlier stage of the
amplitude-computation pipeline. We leave these avenues for future
work.

\section*{Acknowledgements}

We would like to thank the DiRAC and NVIDIA staff for the organisation
of the November 2024 Hackathon at Durham University, which allowed us
to optimise our code on the Bede and Tursa clusters. GDL thanks Daniel
Ma\^itre for guidance in the early stages of this work and comments on
the manuscript, Ben Page for helpful discussions, and the Particle
Theory Group at the Paul Scherrer Institute (PSI) for access to the
gMerlin 6 and 7 clusters. GDL's work is supported in part by the
U.K.\ Royal Society through Grant URF\textbackslash R1\textbackslash
20109.

\newpage

\appendix

\section{CUDA Meta-Programming with Python}\label{app:metaprog}

A central feature of the implementation of \textsc{Linac} is the use
of lightweight code generation to produce specialised CUDA kernels at
runtime directly from Python. Rather than compiling a single generic
kernel capable of handling arbitrary matrix dimensions, arithmetic
domains, and field characteristics, the library dynamically generates
CUDA source code with these quantities promoted to compile-time
constants. This enables more aggressive compiler optimisation than
would otherwise be possible.

The core helper routine responsible for this behaviour is
\texttt{cuda\_set\_vars\_and\_get\_funcs}. It reads a CUDA source
template, substitutes placeholders with runtime parameters, compiles
the resulting source using \textsc{PyCUDA}, and exposes the compiled
CUDA kernels back to Python. In simplified form,
\vspace{-3mm}
\begin{minted}[frame=lines,framesep=2mm,breaklines]{python}
def cuda_set_vars_and_get_funcs(path_to_cuda_script=None, **kwargs):

    oFile = open(path_to_cuda_script, "r")
    sFile = oFile.read()

    for var_name in kwargs:
        sFile = sFile.replace("{" + var_name + "}",
                              str(kwargs[var_name]))

    sCommand = f"""
mod = SourceModule(str(\"\"\"{sFile}\"\"\"))
"""
\end{minted}
\vspace{-1.5mm}
The CUDA source files therefore contain symbolic placeholders such as
\vspace{-3mm}
\begin{minted}[frame=lines,framesep=2mm,breaklines]{c}
#define FIELD_CHARACTERISTIC {FIELD_CHARACTERISTIC}
#define NBR_ROWS {NBR_ROWS}
#define NBR_COLUMNS {NBR_COLUMNS}
\end{minted}
\vspace{-1.5mm}
which are replaced at runtime before compilation.

This approach allows arithmetic properties of the problem to be known
statically by the compiler. In particular, the field characteristic
\(p\) becomes a compile-time constant, enabling modular arithmetic to
be optimised into specialised low-level integer operations tailored to
the specific modulus. Similarly, fixed matrix dimensions permit
simplifications in index arithmetic and loop structure.

The helper routine also parses the CUDA source in order to expose
compiled global CUDA functions to Python automatically. It scans the
section of the source labelled \texttt{GLOBAL FUNCTIONS} and generates
the corresponding \texttt{mod.get\_function(...)} calls. This avoids
having to maintain a separate list of Python bindings for each CUDA
kernel.

Finally, several launch parameters are chosen at the Python level
before kernel invocation. For example, thread-block sizes are rounded
to multiples of the CUDA warp size,
\texttt{round\_to\_multiple\_of(..., 32)} while helper functions such
as \texttt{number\_of\_foldings} and
\texttt{folded\_number\_of\_columns} are used to partition large
matrices into appropriately sized blocks.

In practice, the resulting compilation overhead is typically only of
order \(\mathcal{O}(100\,\mathrm{ms})\), and is therefore subdominant
for the large reductions targeted by \textsc{Linac}. For the present
applications, the advantages of runtime specialisation and
meta-programming substantially outweigh this additional overhead.

\section{Self-Hosted GitHub Actions Runner with GPU Access}
\label{app:github-actions-runner}

Continuous integration for GPU software requires access to suitable
hardware during testing. While GitHub Actions provides a convenient
framework for automated testing, standard runners are not generally
sufficient for testing CUDA code on the target hardware. To support
the development of \textsc{Linac}, we therefore use a lightweight
self-hosted GitHub Actions runner with GPU access through Docker.

The setup is available at
\begin{center} \href{https://github.com/GDeLaurentis/docker-gpu-runner-for-github-actions}{\texttt{github.com/GDeLaurentis/docker-gpu-runner-for-github-actions}}\, .
\end{center}
It provides a minimal Docker-based environment in which a machine with
an NVIDIA GPU can register itself as a temporary self-hosted runner,
execute workflow jobs, and deregister itself at shutdown. The
container is launched with GPU access enabled via
\vspace{-3mm}
\begin{minted}[frame=lines,framesep=2mm,breaklines]{bash}
docker run --rm --name runner --gpus all runner-image
\end{minted}
\vspace{-1.5mm}
after which it connects to GitHub, creates a self-hosted runner for
the configured repository, and starts listening for jobs. The runner
then becomes visible from the repository settings page under
\texttt{Settings/Actions/Runners}.

The Docker image is based on an NVIDIA CUDA development environment,
so that tools such as \texttt{nvcc} and \texttt{nvidia-smi} are
available inside the container. This makes it possible to test the
CUDA-dependent parts of \textsc{Linac}, including compilation through
\textsc{PyCUDA}, directly within the continuous-integration pipeline.
For debugging, the same image can be started interactively by
overriding the container entry point,
\vspace{-3mm}
\begin{minted}[frame=lines,framesep=2mm,breaklines]{bash}
docker run -it --entrypoint /bin/bash --gpus all runner-image
\end{minted}
\vspace{-1.5mm}
which allows one to inspect the runtime environment and verify the
visibility of the CUDA toolchain.

Authentication with GitHub is handled through repository-specific
configuration files containing the relevant access token and runner
registration information. These files are excluded from version
control. The resulting setup is intentionally simple while still
providing a reproducible and isolated environment for GPU-enabled
continuous integration.

This infrastructure is not specific to \textsc{Linac}. It provides a
small, reusable pattern for testing CUDA-based scientific Python
packages on self-hosted hardware, while retaining the convenience of
GitHub Actions for the rest of the development workflow.

\section{Tensor-Valued Functions}
\label{app:tensor-functions}

In many physics applications, one naturally encounters functions whose
values are vectors or tensors rather than scalar quantities. For
example, a single numerical evaluation may return a collection of
coefficients, form factors, or tensor components. To make such objects
convenient to manipulate, \textsc{Linac} provides the utility class
\texttt{tensor\_function}, which represents functions whose values are
arrays.

A callable returning a \texttt{numpy.array} can be wrapped as
\vspace{-3mm}
\begin{minted}[escapeinside=||,mathescape=true,frame=lines,framesep=2mm,breaklines]{python}
F = tensor_function(callable_function)
\end{minted}
\vspace{-1.5mm}
after which array-like operations can be performed at the level of the
function itself.

The most important operation is indexing. If
\[
  F : X \longmapsto F(X) \in \mathbb{F}^{n_1\times\cdots\times n_r},
\]
then \texttt{F[index]} returns a new lazy
\texttt{tensor\_function} representing the function obtained by
applying \texttt{index} to the output of \(F\). The index can be any
index accepted by the object returned by \texttt{F(X)}, typically a
\texttt{numpy.ndarray}. Thus one may select a single component,
\texttt{F[i]}, a slice, \texttt{F[i:j]}, or more general selections
such as \texttt{F[:, pivots]} or \texttt{F[mask]}, provided these are
valid indices for the array returned by the underlying callable. In
this sense, \texttt{tensor\_function} gives a \texttt{numpy}-like
indexing interface to functions returning arrays, rather than to
arrays themselves.

The class also implements flattening through the method
\texttt{F.flatten()}, which returns a one-dimensional
\texttt{tensor\_function}. This makes it possible to iterate over the
components of an array-valued function, or to pass them as a collection
of scalar functions to other routines.

Right multiplication by a numerical matrix is also supported through
\texttt{F @ A}. This represents the function obtained by evaluating
\(F\) and multiplying the result by \(A\) on the right. It is useful,
for instance, when applying a change of basis to a vector of functions.

The shape of a \texttt{tensor\_function} is inferred lazily from the
first evaluation. This avoids requiring symbolic information about the
output tensor in advance, which is useful when the underlying function
is only available numerically. The class is also compatible with a
lightweight memoisation infrastructure, so that repeated evaluations
can be cached when a \texttt{diskcache}\footnote{See
\href{https://github.com/grantjenks/python-diskcache}{github.com/grantjenks/python-diskcache}.}
attribute is available.

Finally, \texttt{tensor\_function} provides a direct bridge to the
function-space abstractions of section~\ref{sec:applications}. The
method
\vspace{-3mm}
\begin{minted}[escapeinside=||,mathescape=true,frame=lines,framesep=2mm,breaklines]{python}
F.as_span(input_generator, field=Field("finite field", 2**31 - 1, 1))
\end{minted}
\vspace{-1.5mm}
constructs the \texttt{VectorSpaceOfFunctions} spanned by the
components of \(F\), using the supplied sampling-point generator. In
this way, array-valued numerical functions can be treated directly as
families of functions whose linear span is to be studied.

\newpage

\bibliographystyle{utphys.bst}
\bibliography{main}

\providecommand{\href}[2]{#2}\begingroup\raggedright\begin{thebibliography}{10}

\bibitem{HEPSoftwareFoundation:2017ggl}
{\bf HEP Software Foundation} Collaboration, J.~Albrecht {\em et al.}, {\em {A
  Roadmap for HEP Software and Computing R{\&}D for the 2020s}}.
  \href{http://dx.doi.org/10.1007/s41781-018-0018-8}{Comput. Softw. Big Sci.
  {\bf 3} (2019) no.~1, 7}, \href{http://arxiv.org/abs/1712.06982}{{\tt
  arXiv:1712.06982 [physics.comp-ph]}}.

\bibitem{HSFPhysicsEventGeneratorWG:2020gxw}
{\bf HSF Physics Event Generator WG} Collaboration, S.~Amoroso {\em et al.},
  {\em {Challenges in Monte Carlo Event Generator Software for High-Luminosity
  LHC}}. \href{http://dx.doi.org/10.1007/s41781-021-00055-1}{Comput. Softw. Big
  Sci. {\bf 5} (2021) no.~1, 12}, \href{http://arxiv.org/abs/2004.13687}{{\tt
  arXiv:2004.13687 [hep-ph]}}.

\bibitem{Kanzaki:2010ym}
J.~Kanzaki, {\em {Monte Carlo integration on GPU}}.
  \href{http://dx.doi.org/10.1140/epjc/s10052-011-1559-8}{Eur. Phys. J. C {\bf
  71} (2011)  1559}, \href{http://arxiv.org/abs/1010.2107}{{\tt arXiv:1010.2107
  [physics.comp-ph]}}.

\bibitem{Roiser:2025bsk}
S.~Roiser, R.~Sch{\"o}fbeck, and Z.~Wettersten, {\em {Rapid event extraction
  and tensorial event adaption: Libraries for efficient access and generic
  reweighting of parton-level events and their implementation in the MadtRex
  module}}. \href{http://arxiv.org/abs/2510.05100}{{\tt arXiv:2510.05100
  [hep-ph]}}.

\bibitem{Seymour:2025fpu}
M.~H. Seymour and S.~Sule, {\em {An NLO-Matched Initial and Final State Parton
  Shower on a GPU}}. \href{http://arxiv.org/abs/2511.19633}{{\tt
  arXiv:2511.19633 [hep-ph]}}.

\bibitem{Bothmann:2021nch}
E.~Bothmann, W.~Giele, S.~Hoeche, J.~Isaacson, and M.~Knobbe, {\em {Many-gluon
  tree amplitudes on modern GPUs: A case study for novel event generators}}.
  \href{http://dx.doi.org/10.21468/SciPostPhysCodeb.3}{SciPost Phys. Codeb.
  {\bf 2022} (2022)  3}, \href{http://arxiv.org/abs/2106.06507}{{\tt
  arXiv:2106.06507 [hep-ph]}}.

\bibitem{Cruz-Martinez:2025kwa}
J.~M. Cruz-Martinez, G.~De~Laurentis, and M.~Pellen, {\em {Accelerating
  Berends{\textendash}Giele recursion for gluons in arbitrary dimensions over
  finite fields}}.
  \href{http://dx.doi.org/10.1140/epjc/s10052-025-14318-3}{Eur. Phys. J. C {\bf
  85} (2025) no.~5, 590}, \href{http://arxiv.org/abs/2502.07060}{{\tt
  arXiv:2502.07060 [hep-ph]}}.

\bibitem{Valassi:2025gmq}
A.~Valassi, {\em {New GPU developments in the Madgraph CUDACPP plugin: kernel
  splitting, helicity streams, cuBLAS color sums}}.
  \href{http://arxiv.org/abs/2510.05392}{{\tt arXiv:2510.05392
  [physics.comp-ph]}}.

\bibitem{Smirnov:2015mct}
A.~V. Smirnov, {\em {FIESTA4: Optimized Feynman integral calculations with GPU
  support}}. \href{http://dx.doi.org/10.1016/j.cpc.2016.03.013}{Comput. Phys.
  Commun. {\bf 204} (2016)  189--199},
  \href{http://arxiv.org/abs/1511.03614}{{\tt arXiv:1511.03614 [hep-ph]}}.

\bibitem{Borowka:2017idc}
S.~Borowka, G.~Heinrich, S.~Jahn, S.~P. Jones, M.~Kerner, J.~Schlenk, and
  T.~Zirke, {\em {pySecDec: A toolbox for the numerical evaluation of
  multi-scale integrals}}.
  \href{http://dx.doi.org/10.1016/j.cpc.2017.09.015}{Comput. Phys. Commun. {\bf
  222} (2018)  313--326}, \href{http://arxiv.org/abs/1703.09692}{{\tt
  arXiv:1703.09692 [hep-ph]}}.

\bibitem{Borowka:2018goh}
S.~Borowka, G.~Heinrich, S.~Jahn, S.~P. Jones, M.~Kerner, and J.~Schlenk, {\em
  {A GPU compatible quasi-Monte Carlo integrator interfaced to pySecDec}}.
  \href{http://dx.doi.org/10.1016/j.cpc.2019.02.015}{Comput. Phys. Commun. {\bf
  240} (2019)  120--137}, \href{http://arxiv.org/abs/1811.11720}{{\tt
  arXiv:1811.11720 [physics.comp-ph]}}.

\bibitem{Heinrich:2021dbf}
G.~Heinrich, S.~Jahn, S.~P. Jones, M.~Kerner, F.~Langer, V.~Magerya,
  A.~P{\"o}ldaru, J.~Schlenk, and E.~Villa, {\em {Expansion by regions with
  pySecDec}}. \href{http://dx.doi.org/10.1016/j.cpc.2021.108267}{Comput. Phys.
  Commun. {\bf 273} (2022)  108267},
  \href{http://arxiv.org/abs/2108.10807}{{\tt arXiv:2108.10807 [hep-ph]}}.

\bibitem{Heinrich:2023til}
G.~Heinrich, S.~P. Jones, M.~Kerner, V.~Magerya, A.~Olsson, and J.~Schlenk,
  {\em {Numerical scattering amplitudes with pySecDec}}.
  \href{http://dx.doi.org/10.1016/j.cpc.2023.108956}{Comput. Phys. Commun. {\bf
  295} (2024)  108956}, \href{http://arxiv.org/abs/2305.19768}{{\tt
  arXiv:2305.19768 [hep-ph]}}.

\bibitem{bib:summit95}
M.~Feldman, {\em New GPU-Accelerated Supercomputers Change the Balance of Power
  on the TOP500}, TOP500 News, 2018.
\newblock
  \url{https://www.top500.org/news/new-gpu-accelerated-supercomputers-change-the-balance-of-power-on-the-top500/}.

\bibitem{DeLaurentis:2020xar}
G.~De~Laurentis, {\em {Numerical techniques for analytical high-multiplicity
  scattering amplitudes}}.
\newblock PhD thesis, Durham U., 2020.

\bibitem{vonManteuffel:2014ixa}
A.~von Manteuffel and R.~M. Schabinger, {\em {A novel approach to integration
  by parts reduction}}.
  \href{http://dx.doi.org/10.1016/j.physletb.2015.03.029}{Phys. Lett. B {\bf
  744} (2015)  101--104}, \href{http://arxiv.org/abs/1406.4513}{{\tt
  arXiv:1406.4513 [hep-ph]}}.

\bibitem{Peraro:2016wsq}
T.~Peraro, {\em {Scattering amplitudes over finite fields and multivariate
  functional reconstruction}}.
  \href{http://dx.doi.org/10.1007/JHEP12(2016)030}{JHEP {\bf 12} (2016)  030},
  \href{http://arxiv.org/abs/1608.01902}{{\tt arXiv:1608.01902 [hep-ph]}}.

\bibitem{Smirnov:2019qkx}
A.~V. Smirnov and F.~S. Chukharev, {\em {FIRE6: Feynman Integral REduction with
  modular arithmetic}}.
  \href{http://dx.doi.org/10.1016/j.cpc.2019.106877}{Comput. Phys. Commun. {\bf
  247} (2020)  106877}, \href{http://arxiv.org/abs/1901.07808}{{\tt
  arXiv:1901.07808 [hep-ph]}}.

\bibitem{Maierhofer:2017gsa}
P.~Maierh{\"o}fer, J.~Usovitsch, and P.~Uwer, {\em {Kira{\textemdash}A Feynman
  integral reduction program}}.
  \href{http://dx.doi.org/10.1016/j.cpc.2018.04.012}{Comput. Phys. Commun. {\bf
  230} (2018)  99--112}, \href{http://arxiv.org/abs/1705.05610}{{\tt
  arXiv:1705.05610 [hep-ph]}}.

\bibitem{Klappert:2020nbg}
J.~Klappert, F.~Lange, P.~Maierh{\"o}fer, and J.~Usovitsch, {\em {Integral
  reduction with Kira 2.0 and finite field methods}}.
  \href{http://dx.doi.org/10.1016/j.cpc.2021.108024}{Comput. Phys. Commun. {\bf
  266} (2021)  108024}, \href{http://arxiv.org/abs/2008.06494}{{\tt
  arXiv:2008.06494 [hep-ph]}}.

\bibitem{Klappert:2019emp}
J.~Klappert and F.~Lange, {\em {Reconstructing rational functions with
  FireFly}}. \href{http://dx.doi.org/10.1016/j.cpc.2019.106951}{Comput. Phys.
  Commun. {\bf 247} (2020)  106951},
  \href{http://arxiv.org/abs/1904.00009}{{\tt arXiv:1904.00009 [cs.SC]}}.

\bibitem{Klappert:2020aqs}
J.~Klappert, S.~Y. Klein, and F.~Lange, {\em {Interpolation of dense and sparse
  rational functions and other improvements in FireFly}}.
  \href{http://dx.doi.org/10.1016/j.cpc.2021.107968}{Comput. Phys. Commun. {\bf
  264} (2021)  107968}, \href{http://arxiv.org/abs/2004.01463}{{\tt
  arXiv:2004.01463 [cs.MS]}}.

\bibitem{Peraro:2019svx}
T.~Peraro, {\em {$\text{FiniteFlow}$: multivariate functional reconstruction
  using finite fields and dataflow graphs}}.
  \href{http://dx.doi.org/10.1007/JHEP07(2019)031}{JHEP {\bf 07} (2019)  031},
  \href{http://arxiv.org/abs/1905.08019}{{\tt arXiv:1905.08019 [hep-ph]}}.

\bibitem{Magerya:2022hvj}
V.~Magerya, {\em {Rational Tracer: a Tool for Faster Rational Function
  Reconstruction}}. \href{http://arxiv.org/abs/2211.03572}{{\tt
  arXiv:2211.03572 [physics.data-an]}}.

\bibitem{Abreu:2020xvt}
S.~Abreu, J.~Dormans, F.~Febres~Cordero, H.~Ita, M.~Kraus, B.~Page, E.~Pascual,
  M.~S. Ruf, and V.~Sotnikov, {\em {Caravel: A C++ framework for the
  computation of multi-loop amplitudes with numerical unitarity}}.
  \href{http://dx.doi.org/10.1016/j.cpc.2021.108069}{Comput. Phys. Commun. {\bf
  267} (2021)  108069}, \href{http://arxiv.org/abs/2009.11957}{{\tt
  arXiv:2009.11957 [hep-ph]}}.

\bibitem{Mangan:2023eeb}
J.~Mangan, {\em {FiniteFieldSolve: Exactly solving large linear systems in
  high-energy theory}}.
  \href{http://dx.doi.org/10.1016/j.cpc.2024.109171}{Comput. Phys. Commun. {\bf
  300} (2024)  109171}, \href{http://arxiv.org/abs/2311.01671}{{\tt
  arXiv:2311.01671 [hep-th]}}.

\bibitem{Hostetter_Galois_2020}
M.~Hostetter, {\em {Galois: A performant NumPy extension for Galois fields}},
  11, 2020.
\newblock \url{https://github.com/mhostetter/galois}.

\bibitem{inproceedingsGaloppo}
N.~Galoppo, N.~Govindaraju, M.~Henson, and D.~Manocha,
  \href{http://dx.doi.org/10.1109/SC.2005.42}{``Lu-gpu: Efficient algorithms
  for solving dense linear systems on graphics hardware.,''} vol.~2005, p.~3.
\newblock 01, 2005.

\bibitem{DeLaurentis:2022otd}
G.~De~Laurentis and B.~Page, {\em {Ans{\"a}tze for scattering amplitudes from
  p-adic numbers and algebraic geometry}}.
  \href{http://dx.doi.org/10.1007/JHEP12(2022)140}{JHEP {\bf 12} (2022)  140},
  \href{http://arxiv.org/abs/2203.04269}{{\tt arXiv:2203.04269 [hep-th]}}.

\bibitem{Chawdhry:2023yyx}
H.~A. Chawdhry, {\em {p-adic reconstruction of rational functions in multiloop
  amplitudes}}. \href{http://dx.doi.org/10.1103/PhysRevD.110.056028}{Phys. Rev.
  D {\bf 110} (2024) no.~5, 056028},
  \href{http://arxiv.org/abs/2312.03672}{{\tt arXiv:2312.03672 [hep-ph]}}.

\bibitem{Abreu:2021asb}
S.~Abreu, F.~Febres~Cordero, H.~Ita, M.~Klinkert, B.~Page, and V.~Sotnikov,
  {\em {Leading-color two-loop amplitudes for four partons and a W boson in
  QCD}}. \href{http://dx.doi.org/10.1007/JHEP04(2022)042}{JHEP {\bf 04} (2022)
  042}, \href{http://arxiv.org/abs/2110.07541}{{\tt arXiv:2110.07541
  [hep-ph]}}.

\bibitem{Cederwall:2025ywy}
M.~Cederwall, J.~Hutomo, S.~M. Kuzenko, K.~Lechner, and D.~P. Sorokin, {\em
  {Some remarks on invariants}}. \href{http://arxiv.org/abs/2509.14350}{{\tt
  arXiv:2509.14350 [hep-th]}}.

\bibitem{harris2020array}
C.~R. Harris, K.~J. Millman, S.~J. van~der Walt, R.~Gommers, P.~Virtanen,
  D.~Cournapeau, E.~Wieser, J.~Taylor, S.~Berg, N.~J. Smith, R.~Kern, M.~Picus,
  S.~Hoyer, M.~H. van Kerkwijk, M.~Brett, A.~Haldane, J.~F. del R{\'{i}}o,
  M.~Wiebe, P.~Peterson, P.~G{\'{e}}rard-Marchant, K.~Sheppard, T.~Reddy,
  W.~Weckesser, H.~Abbasi, C.~Gohlke, and T.~E. Oliphant,
  \href{http://dx.doi.org/10.1038/s41586-020-2649-2}{{\em Array programming
  with {NumPy}}Nature {\bf 585} (Sept., 2020)  357--362}.
  \url{https://doi.org/10.1038/s41586-020-2649-2}.

\bibitem{mpmath}
T.~mpmath~development team, {\em mpmath: a {P}ython library for
  arbitrary-precision floating-point arithmetic (version 1.4.0)}, 2026.
\newblock {\tt https://mpmath.org/}.

\bibitem{giuseppe_de_laurentis_2026_20327732}
G.~D. Laurentis, \href{http://dx.doi.org/10.5281/zenodo.20327732}{{\em
  GDeLaurentis/linac: v1.0.1}}, May, 2026.
\newblock \url{https://doi.org/10.5281/zenodo.20327732}.

\bibitem{giuseppe_de_laurentis_2026_18894183}
G.~D. Laurentis, \href{http://dx.doi.org/10.5281/zenodo.18894183}{{\em
  GDeLaurentis/antares: v0.7.1}}, Mar., 2026.
\newblock \url{https://doi.org/10.5281/zenodo.18894183}.

\bibitem{nvidia-cuda-2025}
{NVIDIA Corporation \& affiliates}, {\em {CUDA} {C}++ {Programming} {Guide}},
  \url{https://docs.nvidia.com/cuda/cuda-c-programming-guide/index.html#}.

\bibitem{DeLaurentis:2023nss}
G.~De~Laurentis, H.~Ita, M.~Klinkert, and V.~Sotnikov, {\em {Double-virtual
  NNLO QCD corrections for five-parton scattering. I. The gluon channel}}.
  \href{http://dx.doi.org/10.1103/PhysRevD.109.094023}{Phys. Rev. D {\bf 109}
  (2024) no.~9, 094023}, \href{http://arxiv.org/abs/2311.10086}{{\tt
  arXiv:2311.10086 [hep-ph]}}.

\bibitem{DeLaurentis:2025dxw}
G.~De~Laurentis, H.~Ita, B.~Page, and V.~Sotnikov, {\em {Compact two-loop QCD
  corrections for Vjj production in proton collisions}}.
  \href{http://dx.doi.org/10.1007/JHEP06(2025)093}{JHEP {\bf 06} (2025)  093},
  \href{http://arxiv.org/abs/2503.10595}{{\tt arXiv:2503.10595 [hep-ph]}}.

\bibitem{DeLaurentis:2026brm}
G.~De~Laurentis, H.~Ita, V.~Kuschke, M.~Ruf, and V.~Sotnikov, {\em {Two-loop
  leading-color QCD corrections for Higgs plus two-jet production in the
  heavy-top limit}}. \href{http://arxiv.org/abs/2605.04009}{{\tt
  arXiv:2605.04009 [hep-ph]}}.

\bibitem{giuseppe_de_laurentis_2026_18881385}
G.~D. Laurentis, \href{http://dx.doi.org/10.5281/zenodo.18881385}{{\em
  GDeLaurentis/syngular: v0.6.0}}, Mar., 2026.
\newblock \url{https://doi.org/10.5281/zenodo.18881385}.

\bibitem{DeLaurentis:2023izi}
G.~De~Laurentis, H.~Ita, and V.~Sotnikov, {\em {Double-virtual NNLO QCD
  corrections for five-parton scattering. II. The quark channels}}.
  \href{http://dx.doi.org/10.1103/PhysRevD.109.094024}{Phys. Rev. D {\bf 109}
  (2024) no.~9, 094024}, \href{http://arxiv.org/abs/2311.18752}{{\tt
  arXiv:2311.18752 [hep-ph]}}.

\bibitem{giuseppe_de_laurentis_2026_20041968}
G.~D. Laurentis, \href{http://dx.doi.org/10.5281/zenodo.20041968}{{\em
  GDeLaurentis/lips: v0.6.1}}, May, 2026.
\newblock \url{https://doi.org/10.5281/zenodo.20041968}.

\bibitem{giuseppe_de_laurentis_2026_18881428}
G.~D. Laurentis, \href{http://dx.doi.org/10.5281/zenodo.18881428}{{\em
  GDeLaurentis/pyadic: v0.3.0}}, Mar., 2026.
\newblock \url{https://doi.org/10.5281/zenodo.18881428}.

\bibitem{DGPS}
W.~Decker, G.-M. Greuel, G.~Pfister, and H.~Sch\"onemann, {\em {\sc Singular}
  {4-4-0} --- {A} computer algebra system for polynomial computations},
  \url{http://www.singular.uni-kl.de}, 2024.

\end{thebibliography}\endgroup
\end{document}